\documentclass[twocolumn,tighten,times]{aastex63}

\usepackage{amsmath}
\usepackage{graphicx}

\newcommand{\be}{\begin{equation}}
\newcommand{\ee}{\end{equation}}

\newcommand\comapeor{COMAP-\emph{EoR}}
\newcommand\comapera{COMAP-\emph{ERA}}
\newcommand\Cii{\hbox{[C\,{\sc ii}]}}

\shorttitle{\comapeor}
\shortauthors{Breysse et al.}

\begin{document}

\title{COMAP Early Science: VII. Prospects for CO Intensity Mapping at Reionization}

\correspondingauthor{Patrick C. Breysse}
\email{pb2555@nyu.edu}

\author[0000-0001-8382-5275]{Patrick C. Breysse}
\affil{Center for Cosmology and Particle Physics, Department of Physics, New York University, 726 Broadway, New York, NY, 10003, U.S.A.}

\author[0000-0003-2618-6504]{Dongwoo T. Chung}
\affil{Canadian Institute for Theoretical Astrophysics, University of Toronto, 60 St. George Street, Toronto, ON M5S 3H8, Canada}
\affiliation{Dunlap Institute for Astronomy and Astrophysics, University of Toronto, 50 St. George Street, Toronto, ON M5S 3H4, Canada}

\author[0000-0002-8214-8265]{Kieran A. Cleary}
\affil{California Institute of Technology, Pasadena, CA 91125, USA}

\author[0000-0003-3420-7766]{H\aa vard T. Ihle}
\affil{Institute of Theoretical Astrophysics, University of Oslo, P.O. Box 1029 Blindern, N-0315 Oslo, Norway}

\author[0000-0002-8800-5740]{Hamsa Padmanabhan}
\affil{Departement de Physique Théorique, Universite de Genève, 24 Quai Ernest-Ansermet, CH-1211 Genève 4, Switzerland}

\author[0000-0003-0209-4816]{Marta B. Silva}
\affil{Institute of Theoretical Astrophysics, University of Oslo, P.O. Box 1029 Blindern, N-0315 Oslo, Norway}

\author[0000-0003-2358-9949 ]{J. Richard Bond}
\affil{Canadian Institute for Theoretical Astrophysics, University of Toronto, 60 St. George Street, Toronto, ON M5S 3H8, Canada}

\author{Jowita Borowska}
\affil{Institute of Theoretical Astrophysics, University of Oslo, P.O. Box 1029 Blindern, N-0315 Oslo, Norway}

\author{Morgan Catha}
\affil{Owens Valley Radio Observatory, California Institute of Technology, Big Pine, CA 93513, USA}

\author{Sarah E. Church}
\affil{Kavli Institute for Particle Astrophysics and Cosmology and Physics Department, Stanford University, Stanford, CA 94305, USA}

\author[0000-0002-5223-8315]{Delaney A.~Dunne}
\affil{California Institute of Technology, Pasadena, CA 91125, USA}

\author[0000-0003-2332-5281]{Hans Kristian Eriksen}
\affil{Institute of Theoretical Astrophysics, University of Oslo, P.O. Box 1029 Blindern, N-0315 Oslo, Norway}

\author[0000-0001-8896-3159]{Marie Kristine Foss}
\affil{Institute of Theoretical Astrophysics, University of Oslo, P.O. Box 1029 Blindern, N-0315 Oslo, Norway}

\author{Todd Gaier}
\affil{Jet Propulsion Laboratory, California Institute of Technology, 4800 Oak Grove Drive, Pasadena, CA 91109}

\author{Joshua Ott Gundersen}
\affil{Department of Physics, University of Miami, 1320 Campo Sano Avenue, Coral Gables, FL 33146, USA}

\author[0000-0001-6159-9174]{Andrew I. Harris}
\affil{Department of Astronomy, University of Maryland, College Park, MD 20742}

\author{Richard Hobbs}
\affil{Owens Valley Radio Observatory, California Institute of Technology, Big Pine, CA 93513, USA}

\author[0000-0001-5211-1958]{Laura Keating}
\affil{Leibniz-Institut für Astrophysik Potsdam, An der Sternwarte 16, 14482 Potsdam, Germany}

\author{James W. Lamb}
\affil{Owens Valley Radio Observatory, California Institute of Technology, Big Pine, CA 93513, USA}

\author{Charles R. Lawrence}
\affil{Jet Propulsion Laboratory, California Institute of Technology, 4800 Oak Grove Drive, Pasadena, CA 91109}

\author{Jonas G. S. Lunde}
\affil{Institute of Theoretical Astrophysics, University of Oslo, P.O. Box 1029 Blindern, N-0315 Oslo, Norway}

\author{Norman Murray}
\affil{Canadian Institute for Theoretical Astrophysics, University of Toronto, 60 St. George Street, Toronto, ON M5S 3H8, Canada}

\author[0000-0001-5213-6231]{Timothy J. Pearson}
\affil{California Institute of Technology, Pasadena, CA 91125, USA}

\author[0000-0001-7612-2379]{Liju Philip}
\affil{Jet Propulsion Laboratory, California Institute of Technology, 4800 Oak Grove Drive, Pasadena, CA 91109}

\author{Maren Rasmussen}
\affil{Institute of Theoretical Astrophysics, University of Oslo, P.O. Box 1029 Blindern, N-0315 Oslo, Norway}

\author{Anthony C. S. Readhead}
\affil{California Institute of Technology, Pasadena, CA 91125, USA}

\author{Thomas J. Rennie}
\affil{Jodrell Bank Centre for Astrophysics, Department of Physics and Astronomy, The University of Manchester, Oxford Road, Manchester, M13 9PL, U.K.}

\author[0000-0001-5301-1377]{Nils-Ole Stutzer}
\affil{Institute of Theoretical Astrophysics, University of Oslo, P.O. Box 1029 Blindern, N-0315 Oslo, Norway}

\author[0000-0003-0545-4872]{Marco P. Viero}
\affil{California Institute of Technology, Pasadena, CA 91125, USA}

\author[0000-0002-5437-6121]{Duncan J. Watts}
\affil{Institute of Theoretical Astrophysics, University of Oslo, P.O. Box 1029 Blindern, N-0315 Oslo, Norway}

\author[0000-0003-3821-7275]{Ingunn Kathrine Wehus}
\affil{Institute of Theoretical Astrophysics, University of Oslo, P.O. Box 1029 Blindern, N-0315 Oslo, Norway}

\author{David P. Woody}
\affil{Owens Valley Radio Observatory, California Institute of Technology, Big Pine, CA 93513, USA}

\collaboration{35}{(COMAP Collaboration)\vspace{-0.56cm}}

\begin{abstract}
We introduce \comapeor, the next generation of the Carbon Monoxide Mapping Array Project aimed at extending CO intensity mapping to the Epoch of Reionization.  \comapeor\  supplements the existing 30~GHz COMAP Pathfinder with two additional 30 GHz instruments and a new 16~GHz receiver.  This combination of frequencies will be able to simultaneously map CO(1--0) and CO(2--1) at reionization redshifts ($z\sim5$--$8$) in addition to providing a significant boost to the $z\sim3$ sensitivity of the Pathfinder.  We examine a set of existing models of the EoR CO signal, and find power spectra spanning several orders of magnitude, highlighting our extreme ignorance about this period of cosmic history and the value of the \comapeor\ measurement.  We carry out the most detailed forecast to date of an intensity mapping cross-correlation, and find that five out of the six models we consider yield signal to noise ratios (S/N) $\gtrsim20$ for \comapeor, with the brightest reaching a S/N above 400.  We show that, for these models, \comapeor\ can make a detailed measurement of the cosmic molecular gas history from $z\sim2$--$8$, as well as probe the population of faint, star-forming galaxies predicted by these models to be undetectable by traditional surveys. 
We show that, for the single model that does not predict numerous faint emitters, a \comapeor-type measurement is required to rule out their existence.  We briefly explore prospects for a third-generation Expanded Reionization Array (\comapera) capable of detecting the faintest models and characterizing the brightest signals in extreme detail.
\vspace{1cm}
\end{abstract}


\section{Introduction}
\label{sec:intro}
The Epoch of Reionization (EoR) remains one of the least-explored periods of cosmic history.  During this final cosmic phase transition, photons from the first luminous sources ionized the intergalactic medium (IGM) for the first time since the emission of the cosmic microwave background (CMB) \citep{Loeb2001,McQuinn2016}.  Observations of the optical depth to the CMB have placed integrated limits on the redshift and duration of reionization \citep{Planck2020}, but the details of the process are still largely unobserved.

The EoR has long been a popular target for line intensity mapping (LIM) survey proposals \citep{Kovetz2017}.  Because they are sensitive to the aggregate emission from all emitting objects at a given redshift, intensity maps are less limited by the extreme faintness of the individual redshift $z\gtrsim 6$ sources that reionized the universe.  By mapping line emission at different observing frequencies and therefore different redshifts, it is in principle possible to map the three-dimensional structure of the Universe as reionization proceeds.

The first target for LIM surveys at the EoR was the 21 cm hyperfine transition from neutral hydrogen \citep{Pritchard2012}.  Several experiments have sought or are seeking to use the 21 cm line to map the gradual disappearance of the neutral IGM across the EoR \citep{Ali2015,Beardsley2016,DeBoer2017}.  The 21 cm line, however, carries little sensitivity to the actual ionizing sources themselves.  Models of star formation and the IGM during the EoR are highly sensitive to assumptions about the interstellar media of star-forming galaxies and extrapolations of the luminosity function of faint, undetectable sources \citep{McQuinn2016}. These properties could be probed by an intensity mapping survey specifically focused on the ionizing sources rather than the IGM.

Intensity mapping using rotational transitions of carbon monoxide (CO), first discussed in \citet{Righi2008} as a possible CMB foreground, traces aggregate emission from the dense molecular gas in which most star formation occurs.  \citet{Lidz2011} demonstrated that such an observation would be particularly powerful during reionization as a complement to 21~cm surveys by probing the formation of new stars thought to provide the bulk of ionizing photons.  A sufficiently high-redshift CO intensity mapping survey would be uniquely able to measure the total abundance of molecular gas during reionization.  Moreover, because LIM measurements are sensitive to the faint end of the galaxy luminosity function, they can determine which galaxies contribute most to that measurement, whether reionization is dominated by rare bright objects or numerous fainter ones.

Realizing this potential for EoR intensity mapping of CO is a major goal of the Carbon Monoxide Mapping Array Project (COMAP).  As described in prior papers in this series \citep{es_I}, the currently-observing COMAP Pathfinder is pursuing CO intensity mapping over three 4 deg$^2$ fields in a frequency band centered at 30~GHz using a 10.4 m antenna at the Owens Valley Radio Observatory (OVRO).  The CO signal in this band is expected to be dominated by CO(1--0) emission from redshifts $z=2.4$--$3.4$, the exploration of which makes up the primary science goals of the Pathfinder \citep{es_V,Li2016}.  These observing frequencies also contain subdominant emission from the CO(2--1) line emitted at $z\sim6$--$8$, spanning a large portion of the EoR.  We introduce here an experimental concept designed to isolate the CO intensity mapping signal from reionization and produce high-quality maps of star-forming molecular gas during that epoch.

\begin{figure}[t!]
\centering
\includegraphics[width=\columnwidth]{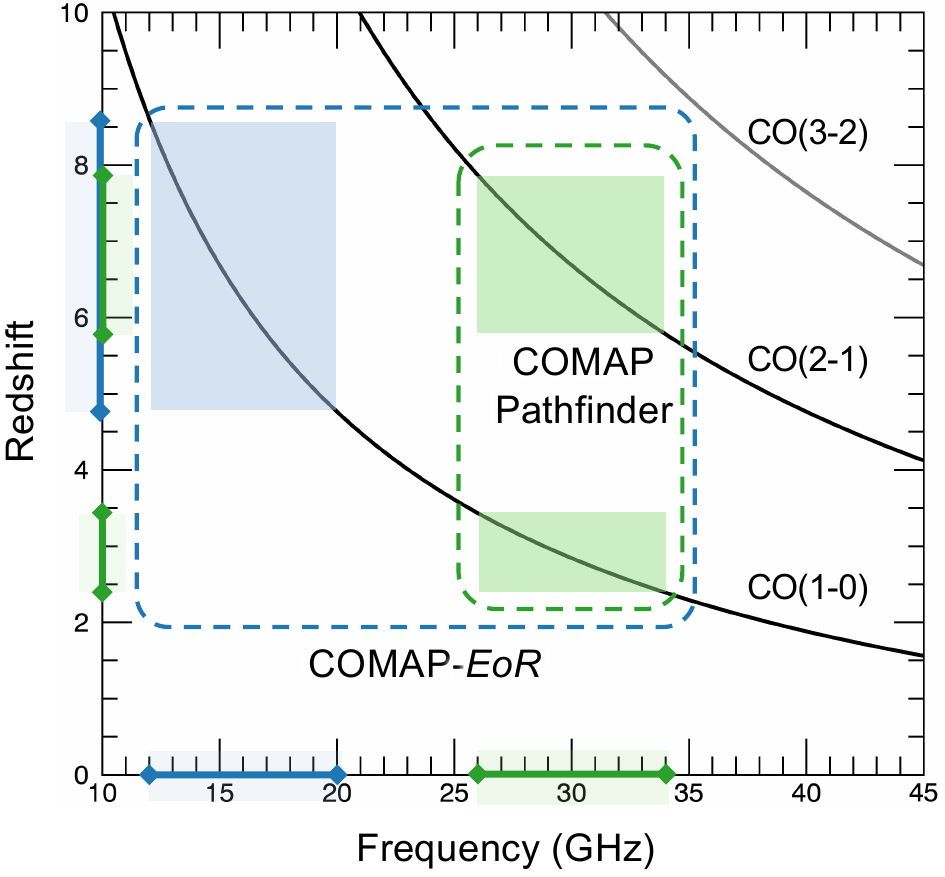}
\caption{Redshift of the three lowest CO transition lines as a function of observed frequency. The frequency coverage for the COMAP Pathfinder Survey (26--34~GHz) is sensitive to the CO(1--0) line in the redshift range $z=2.4$--$3.4$ and the CO(2--1) line at $z=6$--$8$. \comapeor\ adds a second frequency band from 12--20~GHz, sensitive to CO(1--0) from $z=4.8$--$8.6$, allowing a cross-correlation between CO(1--0) and (2--1) during the EoR.
}\label{fig:colines}
\end{figure}

Our planned extension of the existing COMAP Pathfinder, which we term \comapeor, will accomplish this in two ways.  Firstly, more 30~GHz detectors will increase sensitivity to CO(1--0) from $z\sim3$ and CO(2--1) from $z\sim6$--$8$ (during the ``galaxy assembly" and reionization epochs respectively).  Secondly, we will add a second frequency band centered at 16~GHz which will have access to the CO(1--0) transition at the EoR.  We can auto-correlate the 16~GHz maps to measure the EoR signal directly.  We can also cross-correlate the 16 and 30~GHz bands to isolate the EoR CO(2--1) signal from the dominant lower-redshift line.  This will also have the benefit of minimizing any other foreground or systematic effects present in either band.   

In \citet{es_V}, we provided a state-of-the art phenomenological model of CO emission during the $z\sim3$ galaxy assembly epoch which is the focus of the COMAP Pathfinder.  As we will see in this paper, our understanding of reionization remains far too limited to make a similar attempt at $z\sim7$.  The paucity of directly-detected CO emitters at these redshifts means that we cannot fit a useful empirical luminosity function, so we are left with scaling relations and ISM models which are even less certain than they are at $z\sim3$.  Other effects like CMB heating and metallicity evolution further complicate a first-principles modeling effort.  We will instead adopt a method similar to that of \citet{Breysse2014}, and examine the space of existing literature models for the CO signal, using the range of signal amplitudes as a proxy for the detailed uncertainty calculations presented earlier in this series. 
We model the auto-power spectra of CO(1--0) and CO(2--1) at reionization, the galaxy assembly-era CO(1--0) signal which dominates the 30~GHz band, and the cross-spectrum between the two bands.  For all but one of the literature models we predict a highly significant detection of EoR CO emission using the \comapeor\ design.  We go on to show that this measurement can place tight constraints on the abundance of high-redshift molecular gas, and the population of galaxies below the detection threshold of conventional surveys.  We also show a dramatic improvement of the $z\sim3$ CO measurements compared to the Pathfinder forecasts, enabling extremely precise study of molecular gas and star formation in this later epoch.

We also include forecasts for a hypothetical third-generation ``Expanded Reionization Array" stage of COMAP, termed \comapera, intended to follow after the Pathfinder and \comapeor\ surveys.  This survey further increases the sensitivity at both the 16 and 30~GHz bands.  We discuss how this extra depth can allow us to fulfill the promise of tomographic 21 cm and CO cross-correlation, tracing the co-evolution of the interstellar and intergalactic media during the EoR.

For one of our literature models, based on semianalytic simulations presented in \citet{Yang2021sam}, we find a CO signal that is considerably fainter than all of the others, is effectively undetected in \comapeor, and is only seen at the lowest redshifts by \comapera.  We use this model to demonstrate that the CO LIM observations proposed here remain scientifically useful even if only upper limits are obtained. At these sensitivities, our $z\sim3$ CO measurements are quite sensitive to contamination from EoR emission, so directly placing an EoR limit is necessary to reach the potential of the $z\sim3$ measurements, even if no EoR signal is detected.  In addition, a LIM upper limit at $z\sim6$--$8$ in combination with a direct-detection CO survey would serve as proof that there is no significant unseen reservoir of molecular gas during reionization beyond that which is directly imaged.

Section \ref{sec:survey} outlines the experimental design of \comapeor, and we describe our power spectrum formalism in Section \ref{sec:formalism}. Section \ref{sec:models} summarizes the literature models we use for our forecasts, the results of which appear in Section \ref{sec:results}.  We explore the scientific implications of a \comapeor\ detection in Section \ref{sec:science}.  Further discussion appears in Section \ref{sec:discussion}, and we conclude in Section \ref{sec:conclusion}.  Throughout this work, we assume a flat $\Lambda$CDM cosmology consistent with the Planck 2018 results \citep{Planck2020}.  More detail on the COMAP Pathfinder can be found in the other papers in this series, including discussions of the instrumental hardware \citep{es_II}, the data reduction pipeline \citep{es_III}, the power spectrum analysis \citep{es_IV}, the science and modeling implications \citep{es_V}, and the auxiliary Galactic plane observations \citep{es_VI}.

\section{The \comapeor\ Survey}
\label{sec:survey}
Here we will outline the design of the \comapeor\ instrument and survey.  Continued observations with the current 30~GHz Pathfinder instrument will be supplemented with two additional 30~GHz receivers mounted on existing 10.4 m antennas at OVRO along with a new 16~GHz receiver mounted on a new 18 m antenna designed as a prototype for the next-generation Very Large Array (ngVLA). The basic parameters of these instruments can be found in Table \ref{tab:instrument}.  The predicted system temperature for the 30~GHz observations is based on the existing Pathfinder, while that for the 16~GHz instrument is based on expectations for ngVLA given in \citet{Selina2018}.

\begin{deluxetable*}{ccccccccc}
\tablecaption{Parameters of the \comapeor\ instruments, including the central frequency $\nu_{\rm{obs}}$ and width $\Delta\nu$ of each band, the redshift ranges $z_{(1-0)}$ and $z_{(2-1)}$ of each line, the system temperature $T_{\rm{sys}}$, the number of feeds $N_{\rm{feeds}}$ per dish, the beam full width at half maximum $\theta_{\rm{FWHM}}$, and the channel width $\delta\nu$}
\tablehead{
\colhead{Band} & \colhead{Frequency} & \colhead{Bandwidth} & \colhead{CO(1--0) Range} & \colhead{CO(2--1) Range} & \colhead{System Temp.} & \colhead{Feeds/Dish} & \colhead{Beam FWHM} & \colhead{Channel Width} \\
\colhead{ } & \colhead{$\nu_{\rm{obs}}$} & \colhead{$\Delta\nu$} & \colhead{$z_{(1-0)}$} & \colhead{$z_{(2-1)}$} & \colhead{$T_{\rm{sys}}$} & \colhead{$N_{\rm{feeds}}$} & \colhead{$\theta_{\rm{FWHM}}$} & \colhead{$\delta\nu$}}
\startdata
1 & 12.5~GHz & 1~GHz & -- & 7.8 -- 8.6 & 20 K & 38\tablenotemark{a} & 4.2 arcmin & 2 MHz\\
2 & 14~GHz & 2~GHz & -- & 6.7 -- 7.8 & 20 K & 38\tablenotemark{a} & 4.0 arcmin & 2 MHz\\
3 & 16~GHz & 2~GHz & -- & 5.8 -- 6.7 & 22 K & 38\tablenotemark{a} & 3.7 arcmin & 2 MHz\\
4 & 18.5~GHz & 3~GHz & -- & 4.8 -- 5.8 & 27 K & 38\tablenotemark{a} & 3.3 arcmin & 2 MHz\\
\hline
2 & 28~GHz & 4~GHz & 6.7 -- 7.8 & 2.8 -- 3.4 & 44 K\tablenotemark{b} & 19 & 4.5 arcmin & 2 MHz\\
3 & 32~GHz & 4~GHz & 5.8 -- 6.7 & 2.4 -- 2.8 & 44 K\tablenotemark{b} & 19 & 3.9 arcmin & 2 MHz\\
\enddata
\tablenotetext{a}{19 dual-polarization feeds}
\tablenotetext{b}{The current Pathfinder has a system temperature of 44 K, we expect additional instruments to have an improved value of 34 K.  This will be accounted for below in our effective observing time}
\label{tab:instrument}
\end{deluxetable*}

For the early Pathfinder observations reported in other papers in this series, we have assumed a measurement averaged over the entire 8~GHz-wide band of the 30~GHz instrument.  Because the Pathfinder is focused on galaxy-assembly-era measurements near the peak of cosmic star formation, we do not expect the $z\sim3$ CO signal to evolve dramatically over this frequency range, at least at the relatively low sensitivity of the Pathfinder.  For \comapeor, however, we have much more sensitivity to work with, and we have a reionization-era signal which may evolve quite dramatically over our frequency range.  Thus in Table \ref{tab:instrument} and throughout this paper we have divided our observations into four redshift bands.  Two bands centered at 28 and 32~GHz in the high-frequency instruments and 14 and 16~GHz in the low-frequency span the overlapping volume between the CO(1--0) and (2--1) lines at EoR, while two other bands centered at 12.5 and 18.5~GHz account for the additional frequency coverage of the low-frequency instrument.

Given the availability of improved low-noise amplifiers since the deployment of the Pathfinder, we expect the new 30~GHz instruments to have a somewhat lower system temperature, $T_{\rm{sys}}=34$ K compared to the Pathfinder's 44 K.  Below, for ease of forecasting, we will assign $T_{\rm{sys}}=44$ K to all three 30~GHz receivers, and we will scale the effective observing time to account for the improved sensitivity.  The noise level $\sigma_N$ in a map scales as
\be
\sigma_N\propto\frac{T_{\rm{sys}}}{\sqrt{t_{\rm{obs}}}},
\ee
for integration time $t_{\rm{obs}}$, so we can write our total effective observing time as
\be
t_{\rm{obs}}^{\rm{eff}}=t_{\rm{obs}}^{\rm{PF}}+2\left(\frac{44\ \rm{K}}{34\ \rm{K}}\right)^2t_{\rm{obs}}^{\rm{new}},
\label{teff}
\ee
where $t_{\rm{obs}}^{\rm{PF}}$ is the total observing time with the 44 K Pathfinder instrument, and $t_{\rm{obs}}^{\rm{new}}$ is the observing time on each of the new dishes.

For our forecasts here, we will assume that the 16~GHz instrument comes online immediately following the end of the current 5 year Pathfinder campaign, with the two new 30~GHz receivers following an additional two years after that.  The nominal \comapeor\ campaign will then consist of five more years of operation with the full four instruments.  Under this timeline, at the end of those five years we will have accumulated 12 years of time on the Pathfinder, 7 years on the 16~GHz dish, and 5 years on each of the new 30~GHz dishes.  For the \comapeor\ survey, we plan to continue to target the same three fields as the current Pathfinder observations.  Assuming 1000 hours per  year per field of available time, this gives us a total of 29\,000 dish-hours per field at 30~GHz, accounting for the $T_{\rm{sys}}$ adjustment, and 7000 dish-hours per field at 16~GHz.

We will also provide forecasts for a hypothetical third-generation \comapera\ to give an idea of what could be accomplished with even further increases in sensitivity.  As this concept is relatively far in the future, we will model it fairly simply here.  We will assume that, at the end of the above \comapeor\ survey, we increase to 10 dishes at each frequency and observe for an additional five years with all 20.  Assuming these new instruments are identical to the \comapeor\ equivalents, this would give us 110\,000 dish-hours at 30~GHz and 57000 dish-hours at 16~GHz.  For simplicity. here we will assume all of this time is spent on continued observations of the same three Pathfinder fields.  In practice, we may consider other fields in order to cross-correlate with other EoR data, a possibility we will briefly discuss in Section \ref{sec:discussion}.

\section{Power Spectrum Formalism}
\label{sec:formalism}
Because \comapeor\ will observe at 16 and 30~GHz simultaneously, our task in modeling the power spectrum is necessarily more complicated than for the Pathfinder.  We will need to model the reionization-era auto-spectra of the CO(1--0) and (2--1) lines as well as their cross-correlation.  In addition, we require a model of the galaxy assembly-era CO(1--0) auto spectrum.  Though this lower-redshift line is the primary signal for the current Pathfinder observations, it serves as an important foreground to the \comapeor\ CO(2--1) measurement.  Interloper power spectra in intensity mapping surveys become distorted anisotropically when projected into the frame of a higher-redshift signal \citep{Visbal2010,Cheng2016,Lidz2016}, so we will need to model the full angular behavior of the power spectra.  Our formalism here is primarily based on \citet{Bernal2019}, reproduced here for the convenience of the reader.

Throughout this paper, we will make the simplifying assumption that the CO emission in a given dark matter halo comes from a single point source at its center, i.e., we neglect one-halo contributions.  We will also assume that an intensity mapping signal does not evolve across a given frequency band, and will calculate all of our power spectra at a redshift corresponding to the band center.  We also neglect for now the line-broadening effects discussed in \citet{Chung2021}.  As we will see below, we expect any inaccuracies arising due to these assumptions to be small compared to the overall modeling uncertainty.

\subsection{Auto-spectra}

The analytic form for an intensity mapping auto-spectrum $P(k,\mu,z)$ can be written as
\be
P(k,\mu,z)=P^{\rm{clust}}(k,\mu,z)+P^{\rm{shot}}(z),
\ee
where $k$ is the magnitude of the wave vector of a given Fourier mode and $\mu$ is the cosine of the angle between that mode and the line of sight.  

On large scales, the power spectrum is dominated by clustered emission which traces the large-scale structure, that takes the form
\be
P^{\rm{clust}}=\left<Tb\right>^2(z)F^2_{\rm{RSD}}(k,\mu,z)P_m(k,z).
\ee
The shape of the power spectrum is set by the dark matter power spectrum $P_m$, computed here using \texttt{CAMB} \citep{Lewis2002}.  The overall amplitude of the intensity mapping power is set by the luminosity-weighted bias of the target line emitters, given by
\be
\left<Tb\right>=C_{\rm{LT}}(z)\int_{M\rm{min}}^{\infty}L(M,z)b(M,z)f_{\rm{duty}}(M,z)\frac{dn}{dM}dM,
\ee
where $L(M,z)$ is the mean CO luminosity of a halo with mass $M$, $f_{\rm{duty}}(M,z)$ is the fraction of halos with mass $M$ which emit CO at any given time, $b(M,z)$ is the bias for a halo of mass $M$ \citep{Tinker2010}, and $dn/dM$ is the halo mass function, for which we assume the form of \citet{Tinker2008}.  The clustering amplitude $\left<Tb\right>$ is often expressed as a product of the mean line intensity $\left<T\right>$, here expressed in brightness temperature units, and the bias $b$ of the emitting galaxies, which is defined below. In order to ensure that the mass function integral converges, we assume that only halos with masses above $M_{\rm{min}}$ emit CO, with the value of $M_{\rm{min}}$ set by the model under consideration.  The factor
\be
C_{\rm{LT}}(z)=\frac{c^3(1+z)}{8\pi k_B\nu^3H(z)}
\label{clt}
\ee
is the conversion factor between luminosity density and brightness temperature, where $c$ is the speed of light, $k_B$ is Boltzmann's constant, $H(z)$ is the Hubble parameter, and $\nu$ is the rest frequency of the target line.  We can also separate out an average bias factor
\be
b(z)=\frac{\left<Tb\right>}{\left<T\right>}=\frac{\int_{M\rm{min}}^\infty L(M,z)b(M,z)dn/dM dM}{\int_{M\rm{min}} L(M,z)dn/dM dM},
\ee
which gives the degree to which the galaxies are more strongly clustered than the underlying dark matter.

Because intensity maps are made in redshift space, the observed intensity field is distorted anisotropically compared to the true field \citep[see][for a review]{Hamilton1998}.  The effect of these distortions on the power spectrum is encoded in $F_{\rm{RSD}}$, which is given by
\be
F_{\rm{RSD}}(k,\mu,z)=\left(1+\frac{f(z)}{b(z)}\mu^2\right)\frac{1}{1+(k\mu\sigma_{\rm{FoG}})^2/2}.
\ee
The first term gives the linear Kaiser effect \citep{Kaiser1987}, which dominates on large scales with amplitude set by the logarithmic derivative of the growth factor $f(z)$.  The second term describes the fingers-of-God effect due to small-scale peculiar velocities, for which we assume a Lorentzian form with $\sigma_{\rm{FoG}}=7$ Mpc \citep{Bernal2019}.

Because our CO emission is sourced by a population of discrete galaxies rather than a continuous background, our power spectrum also includes a scale-independent shot noise component given by
\begin{multline}
P^{\rm{shot}}(z)=C_{LT}^2(z)\int_{M\rm{min}}^\infty L^2(M,z)f_{\rm{duty}}(M,z) \\ \times\frac{dn}{dM}e^{\left[\sigma_{\rm{sc}}(M,z)\ln(10)\right]^2}dM.
\end{multline}
The exponential factor comes from the fact that some models account for a scatter in line luminosity among halos with a given mass, rather than assigning all halos the mean $L(M,z)$ \citep{Li2016}.  We make the common assumption that this scatter has a lognormal form with standard deviation $\sigma_{\rm{sc}}(M,z)$ in units of dex \citep{Li2016}, which yields the expression above \citep[see discussion in][]{Breysse2021}.

\subsection{Cross-spectrum}
For \comapeor, we also need to model the cross-correlation between reionization-era CO(1--0) and (2--1).  We can again write the cross-power spectrum $P_\times$ between two intensity mapping lines in the form
\be
P_\times(k,\mu,z)=P^{\rm{clust}}_\times(k,\mu,z)+P^{\rm{shot}}_\times(z).
\ee
The basic forms of the clustering and shot-noise terms are derived in the appendix of \citet{Liu2021}.  The cross-clustering term is
\begin{multline}
P^{\rm{clust}}_\times=\left<Tb\right>_1(z)\left<Tb\right>_2(z) \\ \times F_{\rm{RSD},1}(k,\mu,z)F_{\rm{RSD},2}(k,\mu,z)P_m(k,z),
\end{multline}
where subscripts 1 and 2 indicate values computed for CO(1--0) and CO(2--1) respectively.  Since both lines are sourced by the same discrete sources, we also have a cross-shot noise term
\begin{multline}
P^{\rm{shot}}_\times=C_{\rm{LT},1}(z)C_{\rm{LT},2}(z)\int_{M\rm{min}}^\infty L_1(M,z)L_2(M,z) \\ \times f_{\rm{duty}}(M,z)\frac{dn}{dM}dM.
\label{Pshot_x}
\end{multline}
Note that, by not including a scattering factor in Eq. (\ref{Pshot_x}), we have implicitly assumed that any scatter about the mean $L(M)$ distributions is independent for the two lines.  This is likely a fairly conservative assumption for the cross-shot power given what is seen in simulated CO emitting galaxies \citep{Schaan2021,Yang2021}.

\subsection{CO(1--0) ``Interloper"}
The 30~GHz component of the \comapeor\ survey will include a substantial contribution from galaxy assembly-era CO(1--0) emitters.  Formally, this $z\sim3$ emission constitutes a form of foreground contamination to the EoR signal, which will complicate attempts to measure EoR-era \hbox{CO(2--1)}.  However, as seen in other papers in this series, this ``interloper" provides significant science value in its own right.  For our forecasts here, we will therefore simultaneously model the contributions from $z\sim7$ and $z\sim3$ so that our final molecular gas forecasts cover the full accessible redshift range.

In our 30~GHz data, low-redshift CO(1--0) and high-redshift CO(2--1) will be mapped into a common coordinate system.  Thus we must account for projection effects when dealing with these two lines.  Because this paper is primarily (though not entirely) focused on reionization, here we will project the galaxy-assembly line into the higher-redshift coordinate system.  Thus our actual observed power spectrum at 30~GHz will take the form
\be
P(k,\mu)=P_{2}^{\rm{EoR}}(k,\mu)+P_{1,\rm{proj}}^{\rm{GA}}(k,\mu),
\ee
where the notations ``EoR" and ``GA" (for ``Galaxy Assembly") denote $z\sim7$ and $z\sim3$ quantities respectively. The apparent projected CO(1--0) power spectrum can be written as
\be
P_{1,\rm{proj}}^{\rm{GA}}(k_\parallel,k_\perp)=\frac{1}{\alpha_\parallel\alpha_\perp^2} P_1^{\rm{GA}}\left(\frac{k_\parallel}{\alpha_\parallel},\frac{k_\perp}{\alpha_\perp}\right),
\ee
where
\be
\alpha_\parallel=\frac{H(z_{\rm{EoR}})}{H(z_{\rm{GA}})}\frac{1+z_{\rm{GA}}}{1+z_{\rm{EoR}}},
\ee
and
\be
\alpha_\perp=\frac{D_A(z_{\rm{GA}})}{D_A(z_{\rm{EoR}})}
\ee
are the scalings for modes orientated parallel and perpendicular to the line of sight assuming the Hubble parameter $H(z)$ and the comoving angular diameter distance $D_A(z)$.  The original, unprojected $P_{1}^{\rm{GA}}$ auto-spectrum can be computed using an assumed $L(M)$ model in the same manner as the high-redshift lines.

These projection effects mean that, even though we expect the high-redshift CO(2--1) line to be subdominant to the lower-redshift CO(1--0), we can still hope to obtain some information about the CO(2--1) auto-spectrum.  The projection adds extra anisotropy to the projected power spectrum, which can be used to separate the two signals \citep{Lidz2016,Cheng2016}.  We leave for future work a discussion of other methods which could further improve the accuracy of the CO(2--1) auto spectrum, though we note the extensive literature on the matter particularly in the context of \Cii\ intensity mapping \citep[see, e.g.][]{Gong2014,Silva2015,Breysse2015,Yue2015,Sun2018,Cheng2020}.  In particular, COMAP may be able to make use of cross-correlations with Lyman-$\alpha$ emitters from the HETDEX survey \citep{Hill08,Gebhardt21,Hill21} to isolate the two signals \citep{Chung2019hetdex,silva_etal_21,es_V}.  We also neglect for now any other possible contaminating lines.  \citet{Chung2017} demonstrated that, at 30~GHz, CO(1--0) at $z\sim3$ should dominate any other foreground lines.  We do not expect this fact to change for the EoR CO(1--0) line.  

\subsection{Survey Sensitivity}
To finalize our forecasts, we need to model the expected uncertainty on the above power spectra.  Assuming pure white noise, the noise level $\sigma_N$ in a given map voxel takes the form
\be
\sigma_N = \frac{T_{\rm{sys}}}{\sqrt{N_{\rm{feeds}}\,\delta\nu\, t_{\rm{pix}}}},
\ee
where $t_{\rm{pix}}=t_{\rm{obs}}^{\rm{eff}}(\sigma_{\rm{beam}}^2/\Omega_{\rm{field}})$ is the effective observing time per sky pixel.  Note that we are using the effective observing time from Eq. (\ref{teff}), so we are continuing to assume a single effective system temperature.  $N_{\rm{feeds}}$ is the number of feeds in a single dish. Our noise field then has power spectrum
\be
P_N=\sigma_N^2V_{\rm{vox}},
\ee
where $V_{\rm{vox}}$ is the volume of a voxel, which is assumed to be a square $\sigma_{\rm{beam}}=\theta_{\rm{FWHM}}/\sqrt{8\ln 2}$ on a side and a single frequency-channel deep.

We also need to account for the high- and low-$k$ cutoffs in our sensitivity induced by the limited spatial resolution and survey volume.  For a given theoretical power spectrum $P(k,\mu)$, we can construct an observer-space spectrum
\be
\tilde{P}(k,\mu)=W_{\rm{vol}}(k,\mu)W_{\rm{res}}(k,\mu)P(k,\mu),
\ee
where
\begin{multline}
W_{\rm{vol}}=\left(1-\exp\left[-\left(\frac{k}{k_{\perp}^{\rm{min}}}\right)^2(1-\mu^2)\right]\right) \\ \times \left(1-\exp\left[-\left(\frac{k}{k_{\parallel}^{\rm{min}}}\right)^2\mu^2\right]\right),
\end{multline}
and
\be
W_{\rm{res}}=\eta^2\exp\left\{-k^2\left[\sigma_\perp^2(1-\mu^2)+\sigma_\parallel^2\mu^2\right]\right\}
\ee
cut off the measured power at low- and high-$k$ respectively.  We use the forms of $k^{\rm{min}}_\perp$, $k^{\rm{min}}_\parallel$, $\sigma_\perp$, and $\sigma_\parallel$ from \citet{Bernal2019}.  In order to match effects seen in the Pathfinder survey, we have added an additional main beam efficiency factor $\eta$ to account for the loss of power in the main beam.  We adopt the Pathfinder value of $\eta=0.72$ \citep{es_VI} for the 30~GHz instruments, and assume the new 16~GHz instrument will have $\eta\approx1$.  The $W_{\rm{vol}}$ and $W_{\rm{res}}$ values we assume at 30~GHz produce a sensitivity curve in good alignment with the ``optimistic" 5-year Pathfinder forecasts \citep{es_III}.

With our resolution effects in mind, we can now write the errors on our 16 and 30~GHz auto-spectra as
\be
\sigma^2_{16\rm{GHz}}(k,\mu)=\frac{1}{N_{\rm{modes}}(k,\mu)}\left(\tilde{P}_1^{\rm{EoR}}(k,\mu)+P_N^{16\rm{GHz}}\right)^2,
\ee
and
\begin{multline}
\sigma^2_{30\rm{GHz}}(k,\mu)=\frac{1}{N_{\rm{modes}}(k,\mu)} \\ \times \left(\tilde{P}_2^{\rm{EoR}}(k,\mu)+\tilde{P}_{1,\rm{proj}}^{\rm{GA}}(k,\mu)+P_N^{30\rm{GHz}}\right)^2.
\end{multline}
The number of modes $N_{\rm{modes}}$ available in a bin centered at $k,\mu$ is given by
\be
N_{\rm{modes}}(k,\mu)=N_{\rm{field}}\frac{k^2\Delta k\Delta\mu}{8\pi^2}V_{\rm{field}},
\ee
where $V_{\rm{field}}$ is the comoving volume of a single field and $N_{\rm{field}}=3$ accounts for the information from the three COMAP fields.  We can similarly write the error on the cross-spectrum as
\be
\sigma^2_\times=\frac{1}{2}\left(\frac{\tilde{P}^2_\times}{N_{\rm{modes}}}+\sigma_{16\rm{GHz}}\sigma_{30\rm{GHz}}\right).
\ee
Following similar arguments to those in \citet{Bernal2019}, we can after some algebra write the covariances between the auto- and cross-spectra
\begin{multline}
\sigma^2_{\times-16\rm{GHz}}(k,\mu)=\frac{1}{N_{\rm{modes}}}\tilde{P}_\times(k,\mu) \\ \times\left(\tilde{P}_1^{\rm{EoR}}(k,\mu)+P_N^{16\rm{GHz}}\right),
\end{multline}
\begin{multline}
\sigma^2_{\times-30\rm{GHz}}(k,\mu)=\frac{1}{N_{\rm{modes}}}\tilde{P}_\times(k,\mu) \\ \times \left(\tilde{P}_2^{\rm{EoR}}(k,\mu)+\tilde{P}_{1,
\rm{proj}}^{\rm{GA}}(k,\mu),+P_N^{16\rm{GHz}}\right),
\end{multline}
and between the two auto spectra
\be
\sigma^2_{16\rm{GHz}-30\rm{GHz}}=\frac{1}{N_{\rm{modes}}}\tilde{P}_\times^2
\ee

Though we have worked in $(k,\mu)$ coordinates to this point, it is typically not possible to construct a well-defined $\mu$ from a curved-sky observation where the line-of-sight direction changes with telescope pointing.  We will therefore express our sensitivity forecasts in terms of the multipoles of the power spectrum \citep{Yamamoto2006,Chung2019,Bernal2019}.  We can write the multipole $\ell$ of a given power spectrum as
\be
P_\ell(k)=\frac{2\ell+1}{2}\int_{-1}^1 P(k,\mu)\mathcal{L}_\ell(\mu)d\mu,
\ee
where $\mathcal{L}_{\ell}(\mu)$ is the Legendre polynomial of degree $\ell$.  We will consider here the first three multipoles of each power spectrum, which we expect to contain the vast majority of the usable information.  Our data thus consist of the monopoles, quadrupoles, and hexadecapoles of each of the 16 and 30~GHz auto-spectra as well as the cross-spectrum.  We can write the covariance between each of these components as
\begin{multline}
\mathcal{C}^{XY}_{\ell\ell'}(k)=\frac{(2\ell+1)(2\ell'+1)}{2} \\ \times\int_{-1}^1\sigma_X(k,\mu)\sigma_Y(k,\mu)\mathcal{L}_\ell(\mu)\mathcal{L}_{\ell'}(\mu)d\mu.
\label{covmat}
\end{multline}
As an example to clarify this somewhat cumbersome notation, the covariance between the monopole ($\ell=0$) of the 30~GHz spectrum and the hexadecapole ($\ell'=4$) of the cross-spectrum is given by
\begin{multline}
\mathcal{C}^{30\rm{GHz}-\times}_{04}(k)=\frac{9}{2}\int_{-1}^{1}\sigma_{30\rm{GHz}}(k,\mu)\sigma_\times(k,\mu) \\ \times \mathcal{L}_0(\mu)\mathcal{L}_4(\mu)d\mu.
\end{multline}
The signal-to-noise ratio for a given measurement is then
\be
S/N=\left[\sum_{ij}\mathbf{d}^T_i\mathcal{C}_{ij}\mathbf{d}_j\right]^{1/2},
\ee
where $\mathbf{d}_i$ is the data vector constructed from the multipoles of the three power spectra in bins centered at $k_i$ and $\mathcal{C}_{ij}$ is the covariance matrix constructed from the components of Eq. (\ref{covmat}).

\section{Predicting the Reionization Signal}
\label{sec:models}
\begin{figure}
\centering
\includegraphics[width=\columnwidth]{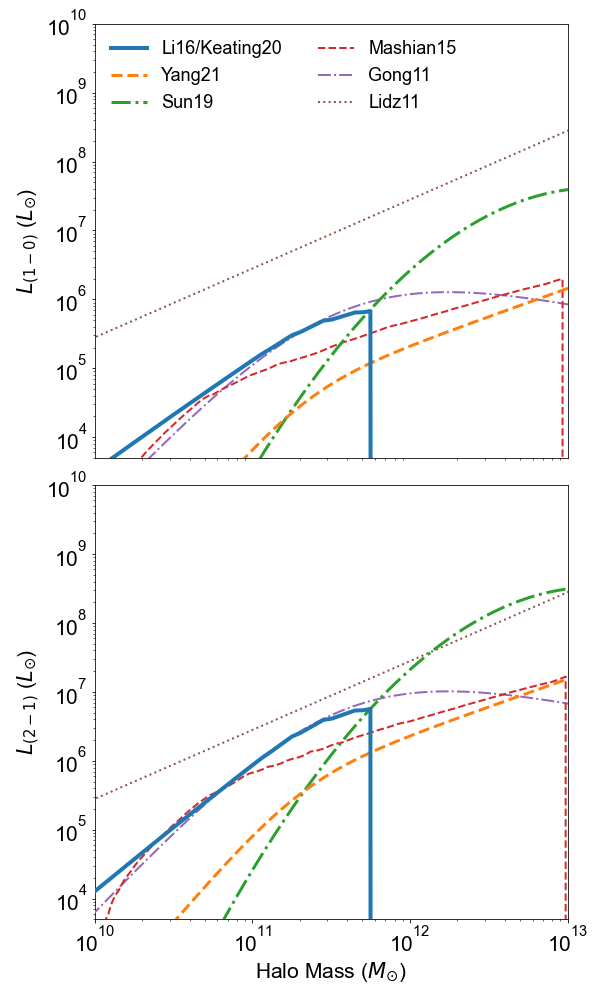}
\caption{CO luminosity $L(M)$ as a function of halo mass for our compilation of literature models, including models from \citet[][brown dotted]{Lidz2011}, \citet[][purple dot-dashed]{Gong2011}, \citet[][red dashed]{Mashian2015}, \citet[][Green dot-dashed]{Sun2019}, \citet[][orange dashed]{Yang2021}, and \citet{Li2016}/\citet[][blue solid]{Keating2020}, with CO(1--0) luminosity plotted in the top panel and CO(2--1) luminosity in the bottom.  All models are computed at $z=6.2$, corresponding to the 16/32~GHz band.}
\label{fig:LofM}
\end{figure}

Here we will briefly summarize the models from the literature that we use to predict the range of possible CO signals at the EoR.   For the full details behind each of these models, please see the relevant references.  When a model does not contain all of the information we need for a full \comapeor\ forecast (for example, only predicting CO(1--0) not CO(2--1)), we will make modest extensions to produce the necessary power spectra.  As these results are intended simply to illustrate the range of possible signal-to-noise ratios (S/N), the exact choice of these extensions should not significantly affect the overall picture.  For our more detailed forecasts in later sections we will confine ourselves to models that inherently provide all of the relevant quantities.   Figure \ref{fig:LofM} compiles the $L(M)$ relations assumed for the CO(1--0) and CO(2--1) lines at reionization.

\subsection{Lidz et al. (2011)}
Most of the models presented here use a method originally put forth by \citet{Lidz2011} to predict CO emission.  Starting from a relationship between star formation rate (SFR) and halo mass, they use empirical scalings between SFR and infrared luminosity $L_{\rm{IR}}$ to get an estimate of $L_{\rm{IR}}(M)$, then use a final empirical scaling to go from $L_{\rm{IR}}$ to $L_{\rm{CO}}$ resulting in the following relation between CO luminosity and halo mass:
\be
\frac{L_{\rm{CO}}(M)}{L_\sun}=2.8\times10^3\left(\frac{M}{10^8\ M_\sun}\right).
\ee
They assume a mass-independent duty cycle $f_{\rm{duty}}=0.1$, do not include any scatter about the mean relation, and assign the same mass-luminosity relation to both CO(1--0) and (2--1).  This latter assumption means that the mean intensity of the CO(2--1) line will be eight times lower than the EoR CO(1--0) due to the factor of $\nu^{-3}$ from Eq. (\ref{clt}).  They explore a number of different $M_{\rm{min}}$ values and we assume their lowest value of $10^8\ M_\sun$ here.  We also assume the same $L(M)$ models for the galaxy assembly and EoR forecasts, though the signal will still evolve through the mass function.  As discussed in \citet{es_V}, this is likely a conservative estimate for this model, at least with regard to the EoR predictions, as the models from \citet{Pullen2013} predict a lower galaxy assembly-era signal using effectively the same method.

\subsection{Gong et al. (2011)}
\citet{Gong2011} make use of a simulated CO catalogue \citep{Obreschkow2009}.  They fit the mean CO mass-luminosity relation with a double power law
\be
L(M)=L_0\left(\frac{M}{M_c}^b\right)\left(1+\frac{M}{M_c}\right)^{-d},
\label{Gongfit}
\ee
where $L_0$, $b$, $M_c$, and $d$ are fit parameters given in their Section 2.  Fit values are given at $z=6$, 7, and 8, we interpolate between them to estimate the fit at other redshifts.  We find that this interpolation gives results in good agreement with their $\left<Tb\right>$ calculations.  For our higher redshift CO(1--0) bin which is centered just above $z=8$, we use their highest redshift values.  \citet{Gong2011} do not assume a duty cycle or a scatter, and they set $M_{\rm{min}}=10^8\ M_\sun$.  Since their forecasts are redshift dependent, but do not extend to $z\sim3$, we assume the below model derived from \citet{Li2016} and \citet{Keating2020} (hereafter Li16/Keating20) for $P_1^{\rm{GA}}$.  As Li16/Keating20 is one of our brighter models, this will be a conservative choice as far as EoR sensitivity is concerned.

As with several of these models, \citet{Gong2011} only make predictions for the CO(1--0) line, so we will need to extrapolate their predictions to CO(2--1).  On the low end, we could follow \citet{Lidz2011} and conservatively assign the same $L(M)$ model to both lines.  On the other hand, \citet{Pullen2013} argue that in the limit of high temperature and optical depth the CO(2--1) mean intensity would be eight times higher than that of CO(1--0).  Given one of these two limits results in $\left<Tb\right>_2=8\left<Tb\right>_1$, and the other gives $\left<Tb\right>_2=\left<Tb\right>_1/8$, we will split the difference here and assume the same mean brightness temperature for the two lines.  This corresponds to increasing the $L(M)$ fit from Eq. (\ref{Gongfit}) by a factor of eight.  This will be our assumption for all models that do not include an explicit CO(2--1) prediction.

\subsection{Mashian et al. (2015)}
\citet{Mashian2015} use a large-velocity gradient (LVG) model \citep{Castor1970,Lucy1971} to predict CO luminosity as a function of halo properties, most notably SFR.  They can then use an abundance-matched estimate of SFR$(M)$ to get their $L(M)$ model.  They assume $f_{\rm{duty}}=1$ and neglect scatter.  We use the ``Photodissociation ON" version of their model, which attempts to account for the destruction of CO molecules due to the radiation background.  We nominally set $M_{\rm{min}}=10^8\ M_\sun$, but this photodissociation effectively cuts off emission below $\sim10^{10}\ M_\sun$.  Models are provided for both CO lines, but only for $z>4$, so we again use the Li16/Keating20 model for galaxy-assembly CO(1--0).

\subsection{Sun et al. (2019)}
\citet{Sun2019} provide models for a number of different intensity mapping lines using a common framework.  They start with an infrared emission model based on cosmic infrared background (CIB) observations \citep{Shang2012,Planck2014}.  They then apply the same mass dependence to the molecular gas mass as a function of halo mass, then transform that into CO luminosity through an assumed $\alpha_{\rm{CO}}$ constant.  They do not apply a duty cycle correction but adopt a $\sigma_{\rm{sc}}=0.3$ dex scatter, and assume $M_{\rm{min}}=10^{10}\ M_\sun$.  As there are only predictions for CO(1--0), we again make the equal-$\left<Tb\right>$ assumption for CO(2--1).  The underlying CIB model is integrated over all redshifts, so we are free to consistently predict both low- and high-redshift CO with this model.

\subsection{Li et al. (2016)/Keating et al. (2020)}
The \citet{Li2016} model formed the basis for the original COMAP Pathfinder forecasts using a more sophisticated version of the \citet{Lidz2011} computation.  CO luminosity is predicted through a chain of scaling relations going through IR luminosity and SFR.  The primary qualitative difference is that \citet{Li2016} made use of abundance-matched SFR-halo mass relations from \citet{Behroozi2013} as opposed to a simple power law.  For their recent millimeter-wave Intensity Mapping Experiment (mmIME) measurements, \citet{Keating2020} applied this same model with newer values for the CO--IR correlations \citep{Kamenetzky2016}, including the addition of higher-$J$ models which we can use to model both of our CO transitions.  The original computation applied scatter in two stages, between halo mass and SFR and between CO and IR luminosity.  At the power spectrum level, however, this was equivalent to assuming a single $\sigma_{\rm{sc}}=0.37$ dex, which we apply here.  Both versions of this model use $f_{\rm{duty}}=1$, and we use their value of $M_{\rm{min}}=10^{10}\ M_{\rm{sun}}$.  Note that the SFR$(M)$ values provided by \citet{Behroozi2013} cut off above $\sim10^{12}\ M_\sun$, so by using these values we effectively assume a maximum halo mass in addition to a minimum.  However, this cutoff exists because halos above that mass are extremely rare in the simulations underlying this model, so we do not expect them to contribute substantially to the LIM signal.

\subsection{Yang et al. (2021)}
Our final CO model, from \citet{Yang2021}, provides fitting functions optimized for intensity mapping based on semianalytic models (SAMs) from \citet{Yang2021sam}.  Unlike most of the above models, which rely heavily on empirical scalings, it attempts to self-consistently model the underlying physics that gives rise to CO emission.  The SAMs are calibrated to a wide variety of galaxy observations, including lower-redshift CO lines.  By providing fitting functions, this model enables easy application of the SAM results to intensity mapping forecasts like our work here.  Mass-luminosity functions here take the form
\be
\frac{L}{L_\sun}=2N\frac{M}{M_\sun}\left[\left(\frac{M}{M_1}\right)^{-\alpha}+\left(\frac{M}{M_1}\right)^\beta\right]^{-1}.
\label{eq:YangLofM}
\ee
The double-power-law shape of Eq.~(\ref{eq:YangLofM}) is common to many data-driven treatments of star formation tracers \citep[see, e.g.][]{Moster2010,Padmanabhan2018}.  Values of $N$, $M_1$, $\alpha$, and $\beta$ are provided for both CO lines, as well as separate fitting functions for $\sigma_{\rm{sc}}$ and $f_{\rm{duty}}$.  We assume $M_{\rm{min}}=10^{10}\ M_\sun$, which here is set by the resolution limits of the semianalytic simulations.  

\section{Sensitivity Forecasts}
\label{sec:results}
\begin{figure*}
\centering
\includegraphics[width=\textwidth]{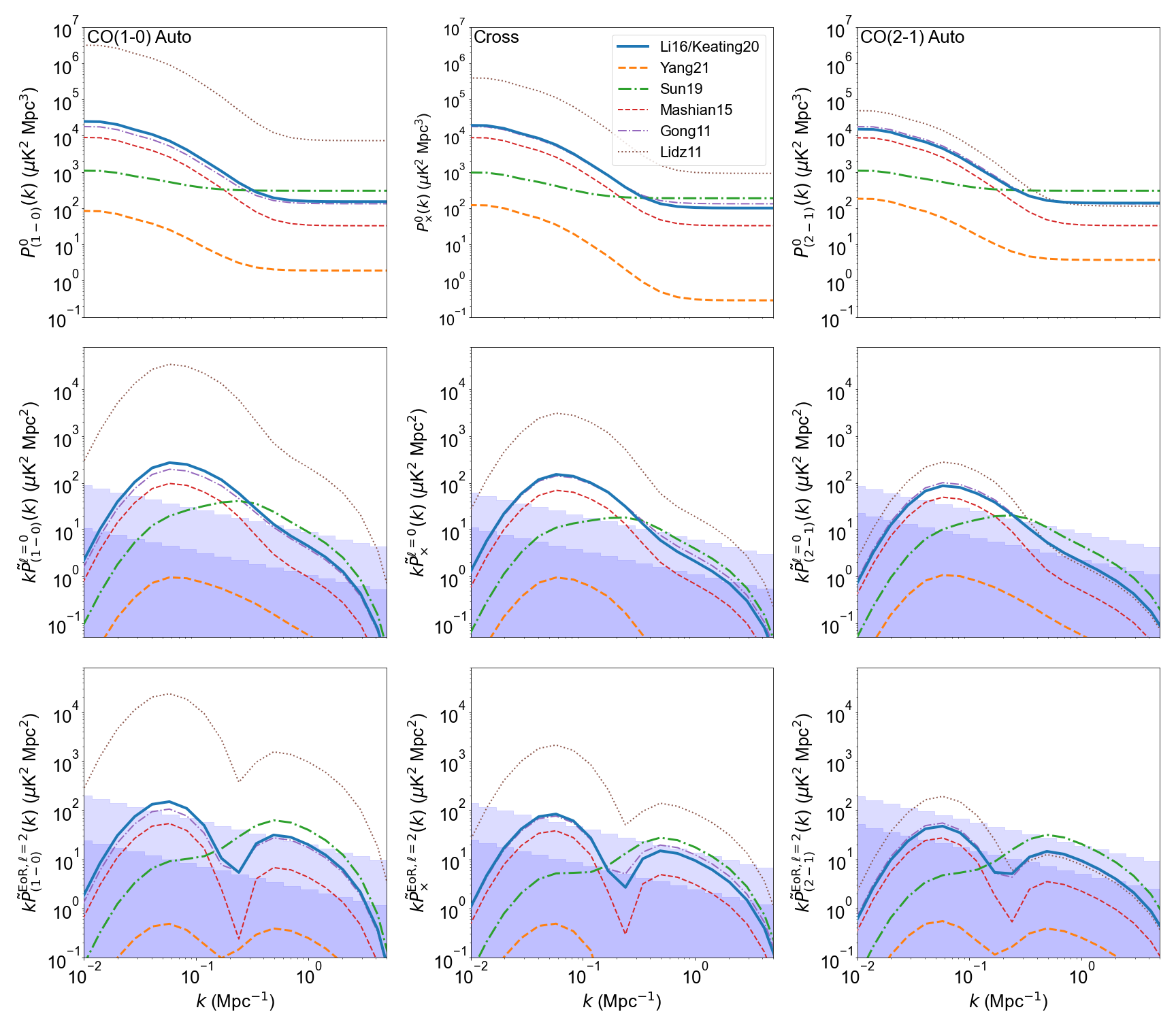}
\caption{Compiled power spectra at $z=6.2$ predicted by the models from Section \ref{sec:models}, using the same color scheme as Fig. \ref{fig:LofM}.  The left column shows the CO(1--0) auto-spectra, the center shows the cross-spectra, and the right column shows the CO(2--1) auto spectra.  The fully theoretical monopole power spectra without observational effects appear in the top row.  The middle row compares the monopole observer-frame spectra (including resolution effects) to the 1-$\sigma$ \comapeor\ and \comapera\ sensitivities, shown here as light and dark shaded bands respectively.  Steps on the sensitivity curves show the assumed $k$ binning.  The bottom row shows the observer-frame quadrupole signals and noise using the same formatting.}
\label{fig:Pkall}
\end{figure*}

We can now apply our power spectrum formalism to each of the above models to see how well the signals they predict can be detected by \comapeor\ and \comapera.  Figure \ref{fig:Pkall} compiles all of the power spectrum forecasts from the above models, in both theoretical and observational form (i.e., with and without resolution effects) compared to sensitivity curves for our two survey concepts.  We can clearly see the primary result of this exercise, that the available models produce an enormous range of possible signals, spanning over four orders of magnitude for CO(1--0).  This highlights the extreme paucity of information about this period of cosmic history.  It also justifies many of the simplifying assumptions made above, as their effect is almost certainly smaller than the range of signals seen here.

Beyond the overall uncertainty, we can see several interesting features in the different models.  The Lidz11 and Yang21 models bracket the range, particularly for CO(1--0) where the extra factor of 8 in the Lidz11 model significantly increases CO(1--0) relative to CO(2--1).  For the other models, even where we have not forced the mean intensities to be identical they still predict similar levels.  Compared to the other models, Sun19 has by far the most shot noise compared to its clustering amplitude due to its quite steep $L(M)$ model, which peaks at higher halo masses than the others.  Finally, the last three models are reasonably consistent with one another, which is perhaps surprising given the overall uncertainty.

\begin{figure}
\centering
\includegraphics[width=\columnwidth]{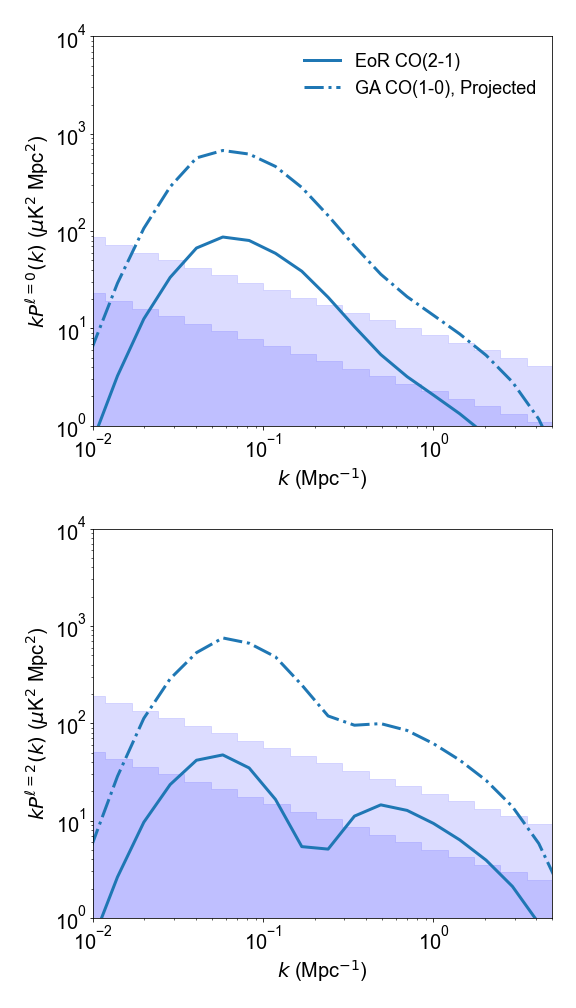}
\caption{Contribution of the projected low-redshift CO(1--0) interloper (dot-dashed) to the reionization-era CO(2--1) (solid) measurement for the Li16/Keating20 model at $z=6.2$.  The monopole spectra are shown in the top panel, the quadrupole spectra in the bottom.  Light and dark shaded regions show the \comapeor\ and \comapera\ sensitivities.}
\label{fig:interloper}
\end{figure}

Figure \ref{fig:Pkall} only shows the EoR power spectra.  As an example of the CO(1--0) ``interloper" effect we compare the CO(2--1) auto-spectrum from the Li16/Keating20 model to the projected galaxy-assembly CO(1--0) in Figure \ref{fig:interloper}.  The effect is relatively small, but we can see by eye that the shapes of the two quadrupole spectra are slightly different, which lets us achieve some modest separation of the two signals.  As shown in the bottom row of Figure \ref{fig:Pkall}, \comapeor\ and particularly \comapera\ should have some quadrupole sensitivity on most of these models, so we should be able to take advantage of this effect.

\begin{deluxetable}{ccccc}
\tablecaption{Signal-to-noise ratios obtained by \comapeor\ and \comapera\ for the auto- and cross-spectra at $z=6.2$, corresponding to $\nu_{\rm{obs}}=16$ and 32~GHz in the low- and high-frequency instruments respectively. Each entry shows the \comapeor\ and \comapera\ S/N's separated by a slash.  Total S/N's combine the significance of the monopole, quadrupole, and hexadecapole measurements.}
\tablehead{
\colhead{Model} & \colhead{CO(1--0)} & \colhead{Cross} & \colhead{CO(2--1)\tablenotemark{a}} & \colhead{Total}
}
\startdata
Li16/Keating20 & 13/54  & 10/31 & 3.4/8.9 & 15/57 \\
Yang21 & 0.1/0.6 & 0.1/0.4 & 0.1/0.2 & 0.1/0.8 \\
Sun19 & 7.5/55 & 6.8/25 & 3.9/13 & 10/60 \\
Mashian15 & 5.1/25 & 7.2/19 & 2.7/7.8 & 7.2/29 \\
Gong11 & 9.9/46 & 14/35 & 1.9/4.9 & 13/53 \\
Lidz11 & 178/455 & 31/52 & 0.2/0.2 & 178/456 \\
\enddata
\label{tab:SNRa}
\tablenotetext{a}{CO(2--1) S/N's include the lower-redshift CO(1--0) power as an extra noise term}
\end{deluxetable}

Table \ref{tab:SNRa} gives the S/N's obtained for \comapeor\ and \comapera\ for the three power spectra in the $z=6.2$ band.  For the CO(2--1) auto S/N, we have effectively treated the interloper as an extra noise component.  For example, this is why the Lidz11 model produces such a low S/N for the CO(2--1) auto-spectrum, as the factor of 8 difference between CO(1--0) and (2--1) means that the low-redshift interloper is much brighter than the EoR line.  This is not precisely correct, as we can perform internal cross-correlations within our raw data to remove any overall bias from instrument noise \citep{es_IV}, while we cannot do the same with a signal on the sky.  However, as mentioned above in a real analysis we could likely do more to mask out the lower-redshift line, which we have not accounted for here.  We thus believe that treating the interloper as noise provides a sufficient approximation for our current level of detail.

\begin{deluxetable}{cccccc}
\label{tab:SNRb}
\tablecaption{S/N's obtained by \comapeor\ and \comapera\ in the four frequency bins from Table \ref{tab:instrument}, along with the combined total.  Values in the two bins which contain both CO(1--0) and CO(2--1) are equivalent to the ``Total" column from Table \ref{tab:SNRa}, the combined sum here assumes the four frequency bins are independent.  Each entry shows the \comapeor\ and \comapera\ S/N's separated by a slash.}
\tablehead{
\colhead{Model} & \colhead{Band 1} & \colhead{2} & \colhead{3} & \colhead{4} & \colhead{Total}
}
\startdata
Li16/Keating20 & 2.2/13 & 9/38 & 15/58 & 21/89 & 27/113 \\
Yang21 & 0.0/0.0 & 0.0/0.1 & 0.1/0.8 & 0.6/4.3 & 0.6/4.4 \\
Sun19 & 0.2/2.0 & 2.3/16 & 10/61 & 22/143 & 24/156 \\
Mashian15 & 0.2/1.2 & 1.9/11 & 7.2/29 & 16/60 & 18/68 \\
Gong11 & 0.3/2.4 & 3.7/20 & 13/53 & 30/115 & 33/128 \\
Lidz11 & 52/126 & 114/290 & 179/456 & 357/811 & 418/983 \\
\enddata
\end{deluxetable}

From Fig.~\ref{fig:Pkall} and Table~\ref{tab:SNRa}, we can see that, in this  redshift bin, \comapeor\ performs quite well for all of the models except that of Yang21.  The Lidz11 model is an outlier in the other direction with very high S/N, while the other models cluster in the S/N $=10$--$20$ range.  Table \ref{tab:SNRb} gives the combined auto+cross S/N's for each of the four frequency bands defined in Table \ref{tab:instrument} (where Band 3 corresponds to the redshift range for Table~\ref{tab:SNRa}).  Assuming there is negligible covariance between the four bands, we then sum the results in quadrature to obtain a total EoR detection significance for each model.

\subsection{Parameter Constraints}
Having studied the overall sensitivity of our planned survey, let us now examine how well we can measure the different components of the power spectrum individually.  We will continue to follow a procedure analogous to \citet{Bernal2019}, but given the very large uncertainty in the galaxy-evolution modeling part of this exercise we will hold all fundamental cosmological parameters fixed.  We have access to three observables (CO(1--0) at galaxy assembly and EoR and CO(2--1) at EoR), and we have a clustering amplitude $\left<Tb\right>$, average bias $b$, and shot noise amplitude $P^{\rm{shot}}$ for each, in addition to the cross-shot power $P^{\rm{shot}}_\times$.  Each of these factors give unique information about the luminosity distribution of the emitting halos.  

We note, we believe for the first time in the literature, that two of these parameters are exactly degenerate even with the benefit of cross-correlation.  Specifically, the CO(2--1) shot power cannot be separated at our current level of detail from the lower-redshift CO(1--0) shot power.  Because Poisson power is scale-independent, we can only measure a single, overall shot amplitude in the 30~GHz power spectrum.  There are not scale-dependent features to shift in the redshift projection, and the shot power only appears in the monopole, so they cannot be separated through the power spectrum anisotropy.  We therefore introduce a new composite quantity
\be
S_{\rm{tot}} = P^{\rm{shot,EoR}}_2+\frac{1}{\alpha_\parallel\alpha_\perp^2}P^{\rm{shot,GA}}_1,
\ee
which is the actual value measurable from the set of power spectra we consider here.

We thus have a total of nine free parameters:
\begin{multline}
\theta_i=\left[\left<Tb\right>_{1}^{\rm{EoR}},\left<Tb\right>_{2}^{\rm{EoR}},\left<Tb\right>_{1}^{\rm{GA}},P^{\rm{shot,EoR}}_{1},S_{\rm{tot}},\right.\\\left.P^{\rm{shot}}_\times,b_{1}^{\rm{EoR}},b_{2}^{\rm{EoR}},b_{1}^{\rm{GA}}\right].
\end{multline}

We will primarily use the Li16/Keating20 model to demonstrate the kinds of parameter constraints and science results we could obtain from \comapeor.  This is a relatively bright (though not the brightest) model, but more importantly it was shown to be broadly consistent with the best existing LIM data for CO(2--1) and above from mmIME \citep{Keating2020}, and perhaps slightly underestimates CO(1--0) LIM observations from COPSS \citep{Keating2016}.  For some of our examples, we will also include predictions from the Yang21 semianalytic model.  As we will see, this model appears to be quite pessimistic even compared to the handful of existing high-redshift CO direct observations, so it provides something of a worst-case scenario for CO at EoR.  These are also the most recently published models from our set, and they both provide all of the information we need for our multi-line, multi-redshift forecasts without any additional assumptions.

We can obtain quantitative parameter forecasts by computing the Fisher matrix,
\be
F_{ij}=\frac{\partial \mathbf{d}^T}{\partial \theta_i}\mathcal{C}^{-1}\frac{\partial\mathbf{d}}{\partial \theta_j},
\ee
where $\mathbf{d}$ and $\mathcal{C}$ are the complete data vector and covariance matrix from Section \ref{sec:formalism}.  The Fisher matrix is then the inverse covariance matrix for the chosen parameters $\theta_i$.  Given the huge modeling uncertainty above, we assume completely flat priors on all parameters.

\begin{figure}
\centering
\includegraphics[width=\columnwidth]{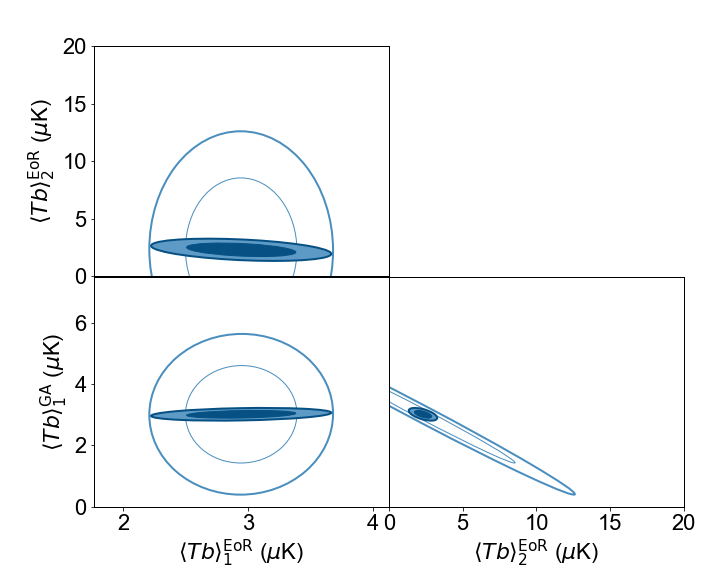}
\caption{Fisher constraints on the power spectrum amplitudes $\left<Tb\right>$ for a \comapeor\ observation at $z=6.2$ assuming the Li16/Keating20 model.  Light and dark filled ellipses show the 95 and 68\% confidence regions for the combination of the two auto-spectra and the cross-spectrum, thin and thick unfilled ellipses show the same but neglecting the cross-correlation.}
\label{fig:LiAmp}
\end{figure}

The full nine-parameter forecast for the Li16/Keating20 model can be found in Appendix \ref{app:Fisher}; for the sake of readability we will highlight some important aspects of it here separately.  As can be seen in Fig. \ref{fig:Pkall}, the COMAP sensitivity is highest in the lower-$k$ clustering regime of the power spectrum \citep[see also][]{es_V}.  Figure \ref{fig:LiAmp} shows the Fisher constraints on the three clustering amplitudes $\left<Tb\right>$ that appear in our measurement.  As expected from the strong overall detection of this model, \comapeor\ obtains a quite strong detection of all three clustering amplitudes.  We have also highlighted the constraints we would obtain in this space if we only had access to the two auto-spectra at 16 and 30~GHz, i.e.\ we neglected $P_{\times}$ in our forecast.  As a result, the EoR CO(2--1) and galaxy-assembly CO(1--0) become quite strongly correlated (though not perfectly correlated, due to the anisotropy).  This illustrates the unique benefit of the cross-correlation ability of the \comapeor\ plan.

\begin{figure}
\centering
\includegraphics[width=\columnwidth]{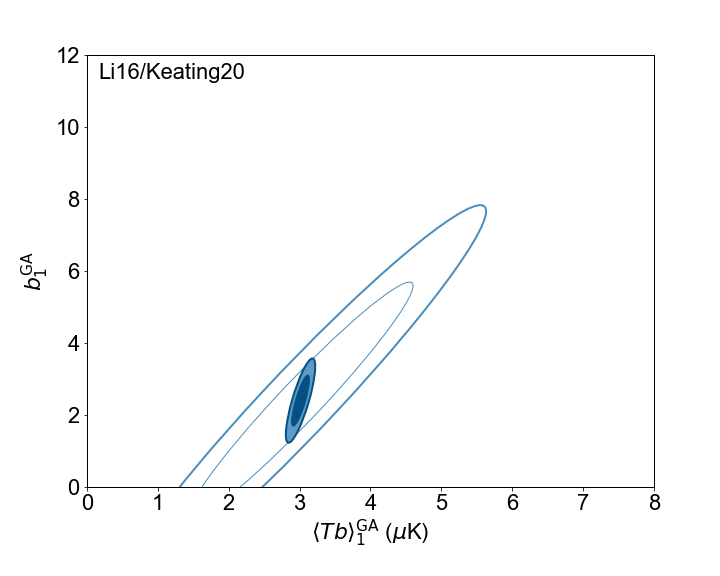}
\includegraphics[width=\columnwidth]{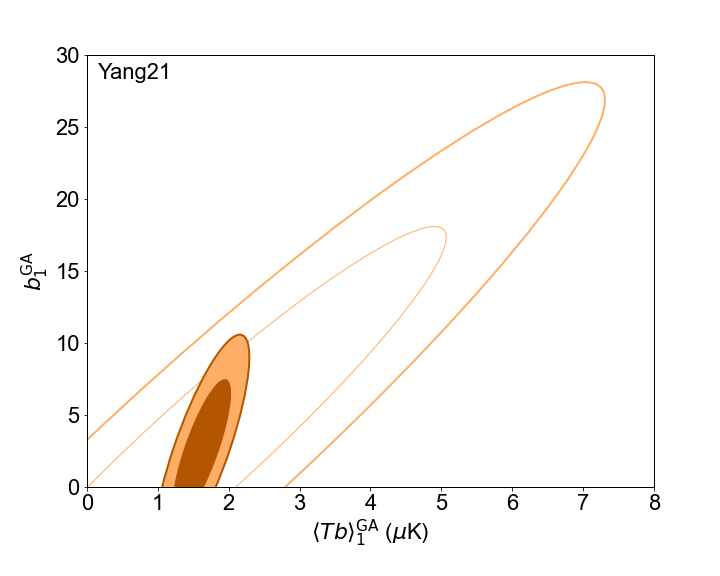}
\caption{Forecasted \comapeor\ constraints on the power spectrum amplitude and bias for the Li16/Keating20 (top) and Yang21 (bottom) models at $z=2.6$.  Filled ellipses include the cross-correlation between the two frequency bands, empty ellipses do not.}
\label{fig:GAFisher}
\end{figure}

Though our focus in this work is primarily on reionization, this cross-correlation has crucial benefits for the lower-redshift science as well, even in the case of the Yang21 model where the EoR signal is undetected.  Figure \ref{fig:GAFisher} shows the constraints on the clustering amplitude and bias of CO(1--0) at galaxy assembly from \comapeor\ for our two demonstration models.  The lower-redshift signal only appears at 30~GHz, so what we have here is effectively the difference between adding and not adding in the 16~GHz instrument.  Since the high-redshift signal is so uncertain, it adds quite a bit of extra error to our attempts to make a precise $z\sim3$ observation.  We have neglected this effect in the other papers in this series, as the current sensitivity is still relatively low, but at \comapeor\ sensitivity this could be a quite substantial effect.  For the brighter Li16/Keating20 model, it makes the difference between separating out the mean intensity and bias and not.  For the Yang21 model, this demonstrates that there are significant science benefits to the 16~GHz observation even if it does not make a strong detection itself.

Though COMAP is in general a clustering-focused measurement, we do see some constraints on the shot powers as well.  The $S_{\rm{tot}}$ combined 30~GHz shot amplitude is quite strongly constrained, though as stated above it cannot be separated into low- and high-redshift components.  We see a $\sim2\sigma$ detection of the CO(1--0) and cross-shot powers at EoR under the Li16/Keating20 model.  \comapeor\ does not have the sensitivity to separate out the mean intensity and bias factors at EoR, but \comapera\ does, at least under this fairly optimistic model.

\section{Science Implications}
\label{sec:science}
Up to this point we have focused on measuring the CO power spectra for our various models.  Now we will move on to examine what these power-spectrum measurements will tell us about the nature of high-redshift galaxies.  We explore two critically important questions: What is the total abundance of star-forming molecular gas during reionization, and what type of galaxies contribute most to that measurement? 

\subsection{Cosmic Molecular Gas Abundance}
One of the primary uses of any CO observation, including this one, is to use the molecule as a proxy for molecular gas, which itself is highly correlated with star formation activity.  We can thus convert our CO power spectrum constraints into a measurement of the cosmic molecular gas history.  We will follow an analogous procedure to what was used in the mmIME observations \citep{Keating2020}.  We assume for now that we precisely know the relationship between CO emission and molecular gas abundance, and that the two are linearly related through the $\alpha_{\rm{CO}}$ parameter.  Quantitatively, we assume that the observed mass-luminosity relation for a given model is given by
\be
L(M,z)=A(z)L_0(M,z),
\ee
where $L_0(M,z)$ is the default model relation used above, and any difference between the predicted and observed CO signal enters through the new parameter $A(z)$.  The cosmic molecular gas abundance is then given by
\be
\rho_{\rm{H2}}(z)=A(z)\alpha_{\rm{CO}}\int_{M\rm{min}}^\infty L'_{\rm{CO(1-0)}}(M,z)\frac{dn}{dM}dM,
\ee
where
\be
\frac{L_{\rm{CO(1-0)}}}{L_\sun}=4.9\times10^{-5}\frac{L'_{\rm{CO(1-0)}}}{\rm{K\ km\ s^{-1}\ pc^2}}
\ee
relates the CO luminosity in physical units to that in the observer units commonly used for $\alpha_{\rm{CO}}$.  We can thus perform a new Fisher forecast for $A(z)$ and easily convert to $\rho_{\rm{H2}}$.  Here we will make the same assumption as mmIME that $\alpha_{\rm{CO}}=3.6\ M_\sun\ (\rm{K\ km\ s^{-1}\ pc^2})^{-1}$.

This is of course a very simplistic way to deal with a highly complex galaxy evolution problem.  \citet{Breysse2021} showed that for mmIME, when carrying out this type of procedure, reasonable changes to the underlying CO model altered the final $\rho_{\rm{H2}}$ result by considerably more than the statistical errors.  The $\alpha_{\rm{CO}}$ scaling alone is far more than a single redshift- and mass-independent constant \citep{Bolatto2013}. However, as we have repeated many times, much of this uncertainty will be swept up in the multiple-order-of-magnitude difference in signal between different models.  In addition, this procedure is qualitatively similar to the assumptions made in common direct-detection CO analyses, where, for example, the choice of $\alpha_{\rm{CO}}$ contributes significant systematic uncertainty \citep[see, e.g.][]{Boogaard2021}

\begin{figure*}
\centering
\includegraphics[width=0.8\textwidth]{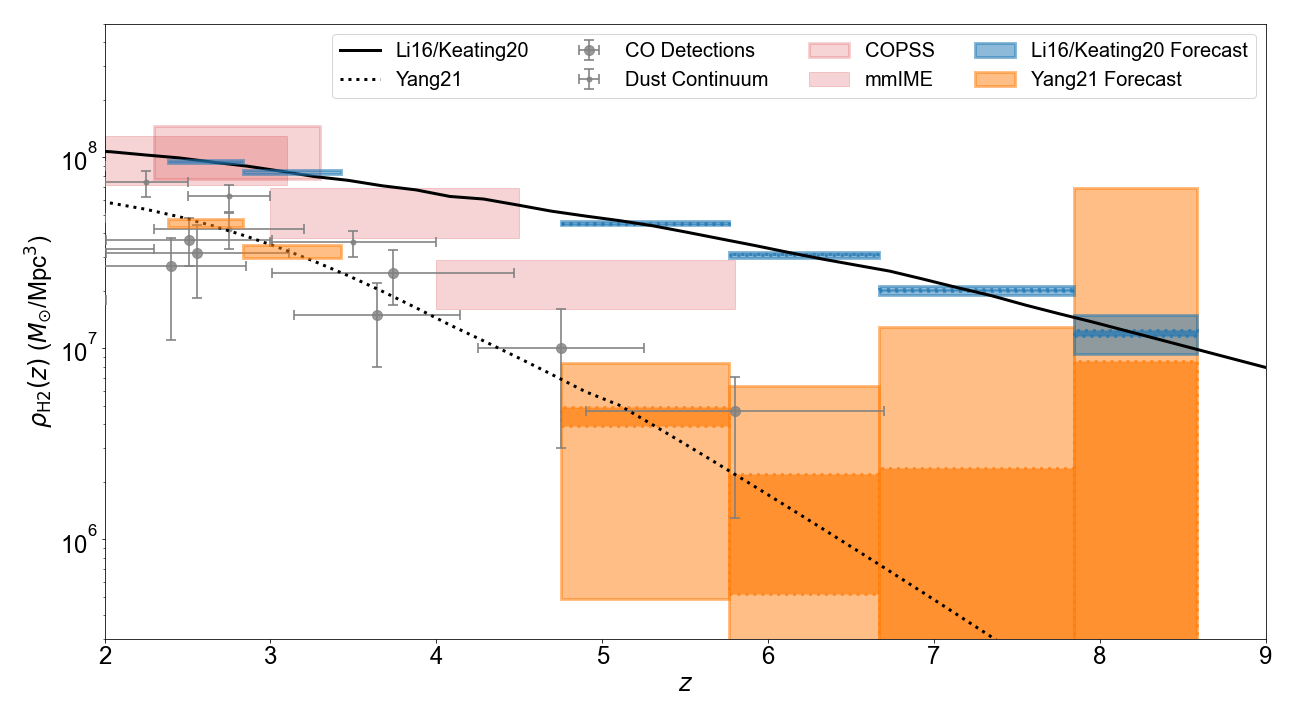}
\caption{Predicted COMAP constraints on the cosmic molecular gas history compared with existing direct and intensity mapping measurements. Gray points and error bars show existing direct observations complied by \citet{Walter2020}, including direct CO observations from ASPECS \citep{2019ApJ...882..136A}, COLDz \citep{2019ApJ...872....7R}, and PHIBBS2 \citep{2020AJ....159..190L}.  Light red boxes show CO intensity mapping constraints from the COPSS (dark) and mmIME (light) surveys.  The solid and dotted black lines show the molecular gas histories inferred from the Li16/Keating20 models assuming a constant $\alpha_{\rm{CO}}=3.6\ M_{\sun}\ (\rm{K}\ \rm{km}/\rm{s}\ \rm{pc}^2)^{-1}$.  Blue and orange boxes show the 95\% constraints obtained on these models using \comapeor\ (light) and \comapera\ (dark).}
\label{fig:rhoH2}
\end{figure*}

Figure \ref{fig:rhoH2} shows forecasts for $\rho_{\rm{H2}}$ measurements from \comapeor\ and \comapera\ assuming the Li16/Keating20 and Yang21 models as a function of redshift.  Forecasts are compared to both existing galaxy surveys using CO and dust observations and to the COPSS and mmIME CO LIM results.  Broadly speaking, the Li16/Keating20 model is most consistent with the $z\sim3$ LIM data, while the Yang21 model is closer to the $z\sim3$ direct data.  It should be noted that the \citet{Yang2021sam} semianalytic model makes its own prediction for $\rho_{\rm{H2}}(z)$ which may differ from what is plotted here, as the semianalytic models provide their own $\alpha_{\rm{CO}}$ values which are allowed to scale with mass and redshift.  Plotted here is what we would infer from the Yang21 CO abundance assuming a constant $\alpha_{\rm{CO}}$.

Comparing the two EoR predictions, we see that the Li16/Keating20 model evolves quite shallowly with redshift, while the Yang21 model falls off even more steeply than the direct observation points.  The baseline \comapeor\ survey provides an extremely tight measurement for the brighter model, but only a weak constraint on the lowest-redshift bin in our worst-case model.  It takes \comapera\ to trace the full evolution of the extremely faint Yang21 case.  Both models at $z\sim3$ perform dramatically better than any current measurement, though it is important to remember that the previously-discussed model dependence is not included in either the current or our forecasted error bars.

\subsection{Faint Star-Forming Galaxies}
\label{sec:FaintGalaxies}
In Fig. \ref{fig:rhoH2}, the Li16/Keating20 model with constant $\alpha_{\rm{CO}}$ implies significantly more molecular gas than has been directly detected to date, in particular compared to the COLDz survey \citep{2019ApJ...872....7R} which provides the highest-redshift data to date.  A similarly flat evolution of the cosmic star formation rate density has been postulated using gamma ray burst counts \citep{Kistler2009}.  In order for that to be the case, there would need to be a significant reservoir of molecular gas present in galaxies too faint to appear in COLDz.  This would in turn have important implications for the nature of reionization, as we would expect quite a bit more ionization-producing star formation activity with corresponding requirements on escape fraction \citep{McQuinn2016}, and that activity would be concentrated in smaller but more numerous sources.    This possibility gets to one of the key motivations for the concept of LIM in general, that it can constrain galaxy populations too faint to observe directly.  In fact, there is already weak evidence at $z\sim3$ for excess star formation appearing in CO and \Cii\ LIM data compared to what we would predict based on galaxy surveys alone (\citet{Breysse2021} and \citet{Yang2021sam} also compare the COPSS and mmIME $\rho_{\rm{H2}}$ values in Fig.~\ref{fig:rhoH2} to direct measurements at the same redshifts).  In this section we will quantify the search for excess faint emission.

\begin{figure}
\centering
\includegraphics[width=\columnwidth]{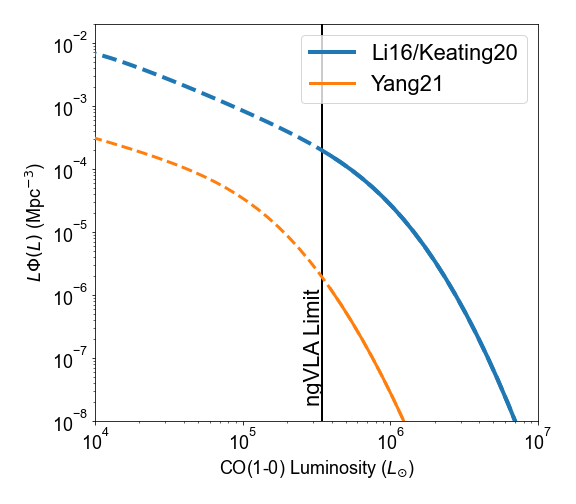}
\caption{CO(1--0) luminosity functions at $z=6.2$ of the Li16/Keating20 (blue) and Yang21 (orange) models, with the limit of the proposed ngVLA molecular gas survey marked in black.  Dashed lines show the portions of the luminosity functions that are directly accessible only to a LIM survey.}
\label{fig:LFs}
\end{figure}

\begin{figure}
\centering
\includegraphics[width=\columnwidth]{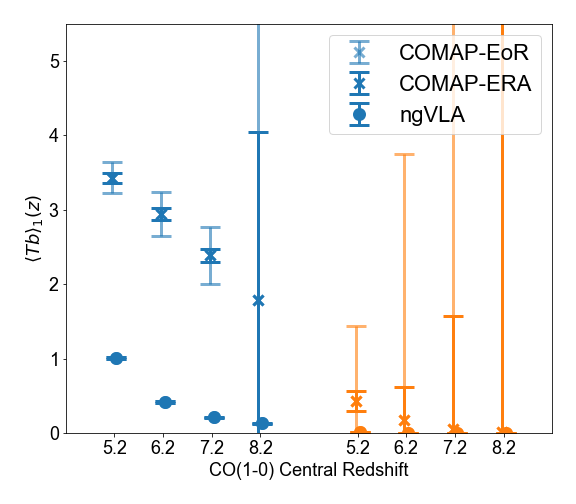}
\caption{Uncertainties on the CO(1--0) power spectrum amplitude factor for the Li16/Keating20 (blue) and Yang21 (orange) models.  Circles show measurements which would be obtained by an ngVLA-like survey, X's show our forecasts for \comapeor\ (light) and \comapera\ (dark).}
\label{fig:FaintGal}
\end{figure}

Figure \ref{fig:LFs} shows the luminosity functions of our two demonstration models in the $z=6.2$ redshift bin.  We can compare these models to the detection limit of a hypothetical CO(1--0) deep field observed with the next-generation Very Large Array (ngVLA), as described in \citet{Decarli2018}.  Then we can ask what is the total CO intensity we would obtain from only those galaxies brighter than the ngVLA limit.  In other words, by comparing the luminosity functions measured from the LIM survey and the direct detection survey we can see how much CO emission (and therefore star formation) is missed in the direct observations.  Figure \ref{fig:FaintGal} shows what happens when we do this.  Since this is a LIM-focused paper, we compute the clustering amplitude $\left<Tb\right>$ for both the entire galaxy population and for only those galaxies brighter than the ngVLA limit.  We then compare the difference between the two to the COMAP errors.  We also make a rough approximation of the error on the ngVLA-determined amplitude assuming Poisson statistics and the 2 deg$^2$ COSMOS-spanning survey described in \citet{Decarli2018}.  For this survey area and the \citet{Decarli2018} detection limits, the Li16/Keating20 model predicts a few thousand CO(1--0) detections, while the Yang21 model predicts a few tens of detections, again highlighting the huge uncertainty in this model space.

For the Li16/Keating20 model, we see that there is a substantial amount of CO being missed by the direct ngVLA survey.  One could always attempt to extrapolate the ngVLA luminosity function to lower values, but only a LIM survey like \comapeor\ could see these faint sources directly.  In the extremely faint Yang21 model, \comapeor\ instead provides a definitive  upper limit on how many faint sources there could possibly be.  Even in the worst-case scenario for EoR detection, this upper limit is still extremely scientifically valuable in ruling out the population seen in Li16/Keating20.

\section{Discussion}
\label{sec:discussion}

Prospects for CO intensity mapping during the EoR are clearly highly dependent on the highly-uncertain signal amplitude.  The current COMAP observing strategy, targeting a handful of relatively small fields, is designed to optimize for the detection of faint signal.  If the true signal is as faint as the Yang21 model predicts, EoR observations will likely remain in this regime up to at least the \comapera\ timescale.  For the rest of the models we consider here, however, there may be motivation in the long term to expand the selected fields to enable more cross-correlation opportunities.

\comapeor\ is primarily designed to map the cross-correlation between CO(1--0) and CO(2--1), but this is far from the only interesting cross-correlation in this redshift range.  As discussed in \citet{silva_etal_21}, a cross-correlation with $z\sim3$ Lyman-$\alpha$ emitters from HETDEX can expand COMAP measurements of both lower-redshift and higher-redshift CO.  \citet{Lidz2011} originally proposed a CO LIM project with the goal of cross-correlating with 21 cm experiments like the Hydrogen Epoch of Reionization Array \citep[HERA;][]{DeBoer2017}.  These two surveys in combination could uniquely measure the typical size of ionized bubbles during the EoR.  Though there are practical difficulties with the differing resolutions between CO and 21 cm observations, the high sensitivity of \comapera\ may make such a measurement possible.

LIM surveys of \Cii\ like the Tomographic Intensity Mapping Experiment \citep[TIME;][]{Crites2014}, the CarbON \Cii\ line in post-rEionization and ReionizaTiOn epoch survey \citep[CONCERTO;][]{Lagache2018}, and the Fred Young Submillimeter Telescope \citep[FYST;][]{CCAT2021} at $z\sim7$ and the Experiment for Cryogenic Large-Aperture Intensity Mapping \citep[EXCLAIM;][]{Cataldo2021} at $z\sim3$ will also map unresolved emission from star-forming galaxies, providing a complement to CO.  Recent literature has proposed intensity mapping of other species at reionization as well, including fine-structure lines of \ion{O}{3} \citep{Padmanabhan2021} and rotational transitions of hydrogen deuteride \citep{Breysse2021hd}.  While true cross-correlations between all of these surveys may be practically difficult, even combining all of these measurements in same model of galaxy evolution will provide powerful insight into the high-redshift ISM.

From our Fisher forecasts in Figures \ref{fig:LiAmp}, \ref{fig:GAFisher}, and \ref{fig:FullFisher}, we can clearly see the benefits of the multi-line cross-correlation approach of \comapeor, advantages which are not so readily available to other LIM targets.  The allowed volume of parameter space reduces dramatically when cross-correlating the two CO lines over the case in which the two auto-spectra are measured separately.  Even when the EoR signal is too faint to detect, it is necessary to actually carry out the observation at both frequencies to make a high-precision measurement of the $z\sim3$ galaxy assembly era.

As mentioned above, however, we have also identified a new limitation of this approach, and indeed all similar LIM cross-correlations.  Because the shot noise in the cross-spectrum has a unique value which cannot be directly determined from the two auto-shot amplitudes, there is no way to separate the shot noise levels between a target line and an interloping foreground line.  This fact may be particularly relevant for the above \Cii\ surveys, as they contain several different CO rotational transitions as interlopers to their EoR signals.  It also may have implications for mmIME-like small-area surveys which are only sensitive to shot power, as this limits their ability to separate out power spectra of different lines.  In both of these cases, individual cross-shot powers will be accessible by cross-correlating pairs of tracers, but at least with cross-correlation and anisotropy alone it will not be possible to isolate any individual auto-shot noise amplitude.  We leave for future work an examination of other interloper-cleaning methods such as voxel masking.

Given the huge range of models shown above and the comparative faintness of the Yang21 model, it is clearly possible that EoR CO emitters will be too faint and rare to be detectable by \comapeor.  Beyond the simple prescriptions discussed here, effects like metallicity evolution and CMB backlighting \citep{daCunha2013} may act to push the signal in the fainter direction, particularly for low-mass galaxies.  Despite this possibility, we argue here that even an upper limit set by a \comapeor-type measurement is still extremely valuable.  All but one of the existing LIM models we consider are bright enough to detect, so we would need a LIM observation to rule them out if nothing else.  As shown in Figure \ref{fig:GAFisher}, a high-redshift upper limit is also critically important to maximizing the science gain from the $z\sim3$ CO maps discussed in the rest of this series.  Finally, even once we have access to high-quality direct observations from ngVLA (which are currently scheduled to appear on the timescale of \comapera, after the nominal \comapeor\ campaign), a LIM upper limit provides confidence that the direct survey has indeed detected the bulk of cosmic star formation during reionization.

The CO power spectra we discuss here do not constitute the entirety of the information available to us in the \comapeor\ data.  Power spectra are two-point statistics, but as shown in \citet{Ihle2019} we can significantly improve a LIM measurement by including the one-point statistics as well, using the Voxel Intensity Distribution (VID) formalism developed in \citet{Breysse2017}.  The power spectra above constrain only the first two moments of the CO luminosity functions, whereas the VID is sensitive to the full distribution.  A joint one- and two-point analysis would improve our ability to constrain more sophisticated models of galaxy evolution, and provide a more detailed description of any faint populations like those shown in Section \ref{sec:FaintGalaxies}.  More recently, \citet{Breysse2019} proposed a conditional one-point formalism which acts as a VID-equivalent for the cross-correlation.  Such a conditional VID estimator or a continuous analogue thereof would allow further non-Gaussian probes of the relationship between the two lines we study here.

\section{Conclusion}
\label{sec:conclusion}
We have presented an overview of the \comapeor\ experiment, the next phase of the COMAP effort.  By adding additional sensitivity at 30~GHz and a new observing band at 16~GHz, we can leverage the ladder of CO rotational transitions to map the CO(1--0) and (2--1) lines in overlapping volumes during the Epoch of Reionization.  By examining CO emission models from the literature, we have shown that the strength of CO emission at $z\sim5-8$ is highly uncertain, a consequence of our poor understanding of the interstellar medium in these early galaxies.  

In order to predict how well \comapeor, and its hypothetical successor \comapera, could detect these models, we carried out the most detailed forecast to date of a cross-correlation between two intensity maps.  For all but one of the models we consider, we predict a highly significant detection of the EoR signal, with $S/N\gtrsim20$.  With such a strong detection, we can isolate the different components of the LIM signal, measuring multiple moments of the two CO luminosity functions.  Using this measurement, we can make a uniquely complete measurement of the total reservoir of star-forming molecular gas during the EoR, and dramatically improve our measurement of the same quantity at $z\sim3$ over what is possible with the COMAP Pathfinder \citep{es_V}.  For the faintest model we consider, \comapeor\ obtains only upper limits on the reionization signal, but we show that even in this worst-case scenario it is still critically important to carry out the measurement in order to remove degeneracies on the $z\sim3$ measurement and to conclusively rule out the possibility of CO emission from galaxies too faint to observe directly.

Under our most pessimistic assumptions, a survey with the sensitivity of \comapera\ will be required to approach a CO detection.  For all of the other above models, however, a $\sim$$100$-$\sigma$ measurement with \comapera\ would open up numerous opportunities for detailed study of the high-$z$ ISM.  At this level, the redshift evolution of the EoR could be followed extremely precisely, and a \citet{Lidz2011}-style cross-correlation between CO and 21 cm may become possible.  More detailed work will be necessary to catalog the possibilities of such a deep LIM observation.

Our extreme ignorance about the CO LIM signal at reionization should serve in its own right as motivation for a survey like \comapeor.  The brightest and faintest models we consider here represent wildly different predictions for the nature of star formation during reionization, and therefore the nature of the EoR itself.  As we have shown, \comapeor\ will radically reduce this uncertainty and open a key new window into this latest cosmic phase transition.

\acknowledgements{
This material is based upon work supported by the National Science Foundation under Grant Nos.\ 1518282, 1910999 and 1517108, and by the Keck Institute for Space Studies under ``The First Billion Years: A Technical Development Program for Spectral Line Observations''.  Parts of the work were carried out at the Jet Propulsion Laboratory, California Institute of Technology, under a contract with the National Aeronautics and Space Administration, and funded through the internal Research and Technology Development program.

PCB is supported by the James Arthur Postdoctoral Fellowship. DTC is supported by a CITA/Dunlap Institute postdoctoral fellowship. The Dunlap Institute is funded through an endowment established by the David Dunlap family and the University of Toronto. HI acknowledges support from the Research Council of Norway through grant 251328. HP acknowledges support from the Swiss National Science Foundation through Ambizione Grant PZ00P2\_179934. Work at the  University of Oslo is supported by the Research Council of Norway through grants 251328 and 274990, and from the European Research  Council (ERC) under the Horizon 2020 Research and Innovation Program (Grant agreement No. 819478, \textsc{Cosmoglobe}). JG acknowledges support from the Keck Institute for Space Science, NSF AST-1517108 and University of Miami and Hugh Medrano for assistance with cryostat design.  JL acknowledges support  from NSF Awards 1518282 and 1910999. LK was supported by the European Union’s Horizon 2020 research and innovation program under the Marie Skłodowska-Curie grant agreement No. 885990.  We thank Isu Ravi for her contributions to the warm electronics and antenna drive characterization.

The authors would like to thank Shengqi Yang, Anthony Pullen, Rachel Somerville, and Abhishek Maniyar for useful discussions.}

\bibliography{references,early_science}{}
\bibliographystyle{aasjournal}

\appendix
\section{Full Fisher Constraints}
\label{app:Fisher}
\begin{figure*}[hb]
\centering
\includegraphics[width=\textwidth]{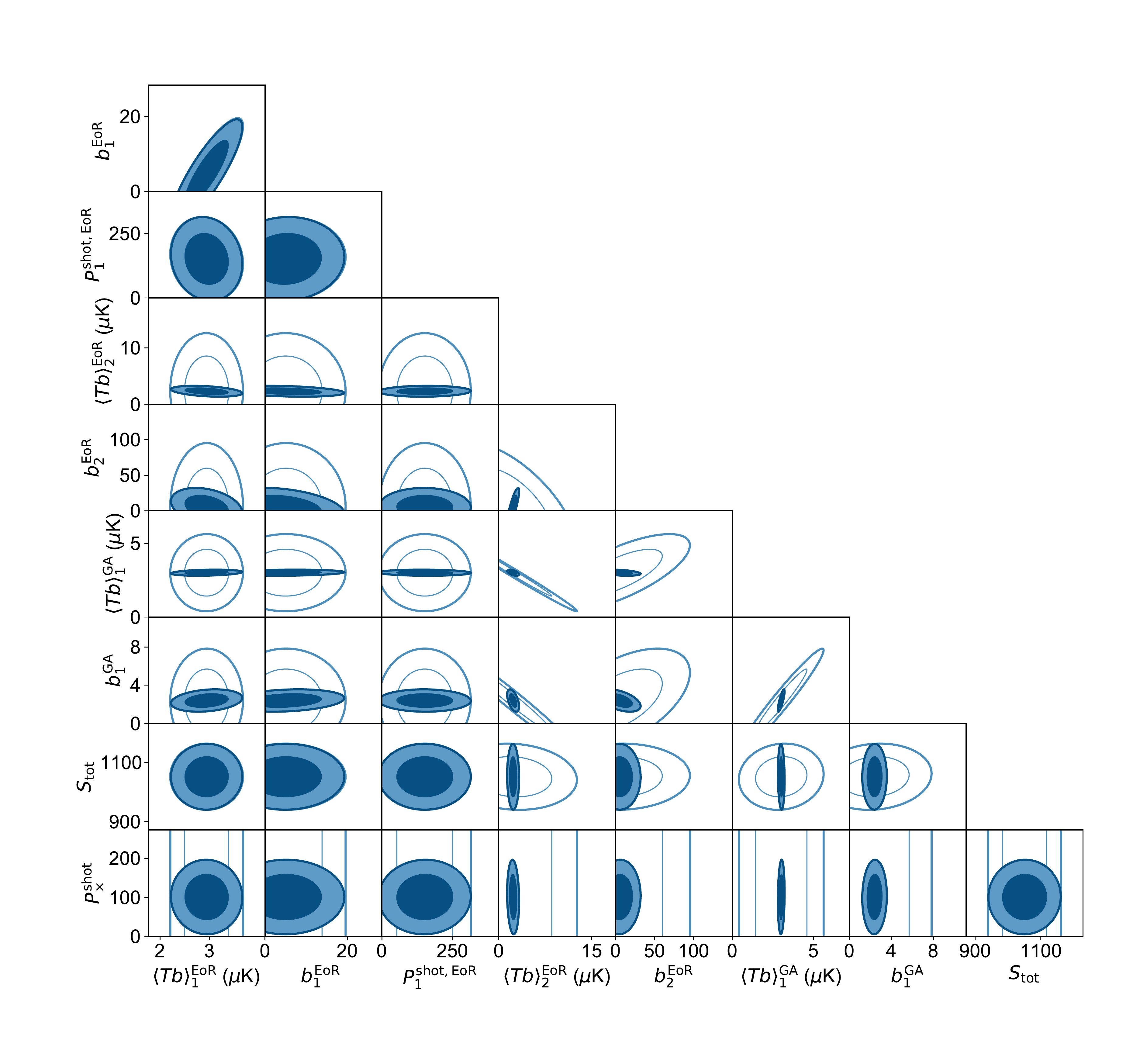}
\caption{Full output of the Li16/Keating20 Fisher forecast.  Light and dark ellipses show the 95 and 68\% confidence regions for the full \comapeor\ observation, thin and thick solid lines show the same for the case where we only use the two auto-spectra and neglect the cross-correlation.}
\label{fig:FullFisher}
\end{figure*}

Figure \ref{fig:FullFisher} shows the full nine-parameter Fisher matrix output for the Li16/Keating20 model at $z=6.2$.  Results for the worst-case Yang21 model are qualitatively similar, but with proportionally lower significance.

\end{document}